\documentclass[journal]{IEEEtran}

\usepackage[utf8]{luainputenc}
\usepackage[english]{babel}
\usepackage{tabu}

\usepackage{url}

\usepackage{xcolor}
\usepackage[absolute]{textpos}

\newcommand{\revision}[1]{{\textcolor{black}{#1}}}

\newcommand{\copyrightstatement}{
    \begin{textblock}{15}(0.5,0.3)    
         \noindent
         \centering
         \textblockcolour{white}
         \footnotesize
         \copyright 2022 IEEE. Personal use of this material is permitted.  Permission from IEEE must be obtained for all other uses, in any current or future media, including reprinting/republishing this material for advertising or promotional purposes, creating new collective works, for resale or redistribution to servers or lists, or reuse of any
copyrighted component of this work in other works.
    \end{textblock}
}

\ifCLASSINFOpdf
   \usepackage[pdftex]{graphicx}
   \graphicspath{{../pdf/}{../jpeg/}{../gfx/}}
\else
\fi
\usepackage{amsmath}
\usepackage{amsfonts}
\usepackage{amssymb}

\DeclareMathOperator*{\argmin}{arg\,min}
\ifCLASSOPTIONcompsoc
  \usepackage[caption=false,font=normalsize,labelfont=sf,textfont=sf]{subfig}
\else
  \usepackage[caption=false,font=footnotesize]{subfig}
\fi
\usepackage{url}

\usepackage{fixltx2e}

\usepackage{cite}

\usepackage{tikz,pgfplots}
\usepackage{pgfgantt}
\usepackage{pdflscape}
\pgfplotsset{compat=newest} 
\usetikzlibrary{shapes, arrows, calc, patterns}
\usetikzlibrary{arrows, calc, positioning, decorations.markings}

\newcommand{\dsplinewidth}{0.25mm}           
\newcommand{\dspblocklinewidth}{0.3mm}       
\newcommand{\dspoperatordiameter}{4mm}       
\newcommand{\dspoperatorlabelspacing}{2mm}   
\newcommand{\dspnoderadius}{1mm}             
\newcommand{\dspsquareblocksize}{8mm}        
\newcommand{\dspfilterwidth}{14mm}           

\pgfarrowsdeclare{dsparrow}{dsparrow}
{
	\arrowsize=0.25pt
	\advance\arrowsize by .5\pgflinewidth
	\pgfarrowsleftextend{-4\arrowsize}
	\pgfarrowsrightextend{4\arrowsize}
}
{
	\arrowsize=0.25pt
	\advance\arrowsize by .5\pgflinewidth
	\pgfsetdash{}{0pt} 
	\pgfsetmiterjoin	 
	\pgfsetbuttcap		 
	\pgfpathmoveto{\pgfpoint{-4\arrowsize}{2.5\arrowsize}}
	\pgfpathlineto{\pgfpoint{4\arrowsize}{0pt}}
	\pgfpathlineto{\pgfpoint{-4\arrowsize}{-2.5\arrowsize}}
	\pgfpathclose
	\pgfusepathqfill
}

\pgfarrowsdeclare{dsparrowmid}{dsparrowmid}
{
	\arrowsize=0.25pt
	\advance\arrowsize by .5\pgflinewidth
	\pgfarrowsleftextend{-4\arrowsize}
	\pgfarrowsrightextend{4\arrowsize}
}
{
	\arrowsize=0.25pt
	\advance\arrowsize by .5\pgflinewidth
	\pgfsetdash{}{0pt}
	\pgfsetmiterjoin
	\pgfsetbuttcap
	\pgfpathmoveto{\pgfpoint{0}{2.5\arrowsize}}
	\pgfpathlineto{\pgfpoint{8\arrowsize}{0pt}}
	\pgfpathlineto{\pgfpoint{0}{-2.5\arrowsize}}
	\pgfpathclose
	\pgfusepathqfill
}

\makeatletter

\pgfkeys{/tikz/dsp/label/.initial=above}

\long\def\dspdeclareoperator#1#2{
	\pgfdeclareshape{#1}
	{
		\savedanchor\centerpoint{\pgfpointorigin}
		\savedmacro\label{\def\label{\pgfkeysvalueof{/tikz/dsp/label}}}
	  \saveddimen\radius
	  {
		  \pgfmathsetlength\pgf@xa{\pgfshapeminwidth}
		  \pgfmathsetlength\pgf@ya{\pgfshapeminheight}
	    \ifdim\pgf@xa>\pgf@ya
	      \pgf@x=.5\pgf@xa
	    \else
	      \pgf@x=.5\pgf@ya
	    \fi
	  }
	  
	  \inheritanchor[from={circle}]{center}
	  \inheritanchor[from={circle}]{mid}
	  \inheritanchor[from={circle}]{base}
	  \inheritanchor[from={circle}]{north}
	  \inheritanchor[from={circle}]{south}
	  \inheritanchor[from={circle}]{west}
	  \inheritanchor[from={circle}]{east}
	  \inheritanchor[from={circle}]{mid west}
	  \inheritanchor[from={circle}]{mid east}
	  \inheritanchor[from={circle}]{base west}
	  \inheritanchor[from={circle}]{base east}
	  \inheritanchor[from={circle}]{north west}
	  \inheritanchor[from={circle}]{south west}
	  \inheritanchor[from={circle}]{north east}
	  \inheritanchor[from={circle}]{south east}
	  \inheritanchorborder[from={circle}]
	  
	  \backgroundpath
	  {
	    \pgfpathcircle{\centerpoint}{\radius}
	    
	    #2
	  }
	
	  \anchor{text}
	  {
			\centerpoint
	    \def\templabelabove{above}
	    \def\templabelbelow{below}
	    \def\templabelleft{left}
	    \def\templabelright{right}
	    \pgfutil@tempdima=\dspoperatorlabelspacing
	    \ifx\label\templabelabove
				\advance\pgf@x by -0.5\wd\pgfnodeparttextbox
				\advance\pgf@y by \radius
				\advance\pgf@y by \pgfutil@tempdima
	    \fi
	    \ifx\label\templabelbelow
				\advance\pgf@x by -0.5\wd\pgfnodeparttextbox
				\advance\pgf@y by -\radius
				\advance\pgf@y by -\pgfutil@tempdima
				\advance\pgf@y by -\ht\pgfnodeparttextbox
	    \fi
	    \ifx\label\templabelleft
				\advance\pgf@x by -\radius
				\advance\pgf@x by -\pgfutil@tempdima
				\advance\pgf@x by -\wd\pgfnodeparttextbox
				\advance\pgf@y by -0.5\ht\pgfnodeparttextbox
				\advance\pgf@y by +0.5\dp\pgfnodeparttextbox
	    \fi
	    \ifx\label\templabelright
				\advance\pgf@x by \radius
				\advance\pgf@x by \pgfutil@tempdima
				\advance\pgf@y by -0.5\ht\pgfnodeparttextbox
				\advance\pgf@y by +0.5\dp\pgfnodeparttextbox
	    \fi
	  }
	}
}

\dspdeclareoperator{dspshapecircle}{\pgfusepathqstroke}

\dspdeclareoperator{dspshapecirclefull}{\pgfusepathqfillstroke}

\dspdeclareoperator{dspshapeadder}{
	\pgfutil@tempdima=\radius
	\pgfutil@tempdima=0.55\pgfutil@tempdima
	
	\pgfmoveto{\pgfpointadd{\centerpoint}{\pgfpoint{0pt}{-\pgfutil@tempdima}}}
	\pgflineto{\pgfpointadd{\centerpoint}{\pgfpoint{0pt}{ \pgfutil@tempdima}}}
	
	\pgfmoveto{\pgfpointadd{\centerpoint}{\pgfpoint{-\pgfutil@tempdima}{0pt}}}
	\pgflineto{\pgfpointadd{\centerpoint}{\pgfpoint{ \pgfutil@tempdima}{0pt}}}
	
	\pgfusepathqstroke
}

\dspdeclareoperator{dspshapemixer}{
	\pgfutil@tempdima=\radius
	\pgfutil@tempdima=0.707106781\pgfutil@tempdima
	
	\pgfmoveto{\pgfpointadd{\centerpoint}{\pgfpoint{-\pgfutil@tempdima}{-\pgfutil@tempdima}}}
	\pgflineto{\pgfpointadd{\centerpoint}{\pgfpoint{ \pgfutil@tempdima}{ \pgfutil@tempdima}}}
	
	\pgfmoveto{\pgfpointadd{\centerpoint}{\pgfpoint{-\pgfutil@tempdima}{ \pgfutil@tempdima}}}
	\pgflineto{\pgfpointadd{\centerpoint}{\pgfpoint{ \pgfutil@tempdima}{-\pgfutil@tempdima}}}
	
	\pgfusepathqstroke
}

\makeatother

\tikzset{dspadder/.style={shape=dspshapeadder,line cap=rect,line join=rect,
	line width=\dspblocklinewidth,minimum size=\dspoperatordiameter}}
\tikzset{dspmultiplier/.style={shape=dspshapecircle,line cap=rect,line join=rect,
	line width=\dspblocklinewidth,minimum size=\dspoperatordiameter}}
\tikzset{dspmixer/.style={shape=dspshapemixer,line cap=rect,line join=rect,
	line width=\dspblocklinewidth,minimum size=\dspoperatordiameter}}

\tikzset{dspnodeopen/.style={shape=dspshapecircle,line width=\dsplinewidth,minimum size=\dspnoderadius}}
\tikzset{dspnodefull/.style={shape=dspshapecirclefull,line width=\dsplinewidth,fill,minimum size=\dspnoderadius}}

\tikzset{dspsquare/.style={shape=rectangle,draw,align=center,text depth=0.3em,text height=1em,inner sep=0pt,
	line cap=round,line join=round,line width=\dsplinewidth,minimum size=\dspsquareblocksize}}
\tikzset{dspfilter/.style={shape=rectangle,draw,align=center,text depth=0.3em,text height=1em,inner sep=0pt,
	line cap=round,line join=round,line width=\dsplinewidth,minimum height=\dspsquareblocksize,minimum width=\dspfilterwidth}}

\tikzset{dspline/.style={line width=\dsplinewidth},line cap=round,line join=round}
\tikzset{dspconn/.style={->,>=dsparrow,line width=\dsplinewidth},line cap=round,line join=round}
\tikzset{dspflow/.style={line width=\dsplinewidth,line cap=round,line join=round,
  decoration={markings,mark=at position 0.5 with {\arrow{dsparrowmid}}},postaction={decorate}}}

\pgfplotsset{
	SISYstyle/.style={
		axis x line=bottom,
		axis y line=middle,
		every axis x label/.style={at={(current axis.above of origin)},anchor=east},
		every axis y label/.style={at={(current axis.above origin)},anchor=south},
		every tick/.style={black,thick},
		/pgf/number format/.cd,
		use comma
	}
}
\newlength\figurewidth
\newlength\figureheight
\begin{document}%
\title{\revision{Rate-Distortion Optimal Transform Coefficient Selection for Unoccupied Regions in Video-Based Point Cloud Compression}}

\copyrightstatement

\author{Christian Herglotz, Nils Genser, and André Kaup %
\thanks{C. Herglotz, N. Genser, and A. Kaup are with the Chair of Multimedia Communications and Signal Processing at the Friedrich-Alexander University Erlangen-N\"urnberg, Germany. Email: \{christian.herglotz, nils.genser, andre.kaup\}@fau.de}}
\maketitle

\begin{abstract}
This paper presents a novel method to determine rate-distortion optimized transform coefficients for efficient compression of videos generated from point clouds. The method exploits a generalized frequency selective extrapolation approach that iteratively determines rate-distortion-optimized coefficients for all basis functions of two-dimensional discrete cosine and sine transforms. The method is applied to blocks containing both occupied and unoccupied pixels in video based point cloud compression for HEVC encoding. In the proposed algorithm, only the values of the transform coefficients are changed such that resulting bit streams are compliant to the V-PCC standard. 
For all-intra coded point clouds, bitrate savings of more than $\mathbf{4}\boldsymbol{\%}$ for geometry and more than $\mathbf{6}\boldsymbol{\%}$ for texture error metrics with respect to standard encoding can be observed. 
 These savings are more than twice as high as savings obtained using competing methods from literature. 
 In the randomaccess case, our proposed method 
outperforms competing V-PCC methods by more than $\mathbf{0.5}\boldsymbol{\%}$. 
\end{abstract}

\begin{IEEEkeywords}
Coding, HEVC, unoccupied pixels, inactive regions, point cloud compression, V-PCC
\end{IEEEkeywords}

%
\IEEEpeerreviewmaketitle

\section{Introduction}
\label{sec:intro}


In recent years, due to an increased interest in 3D representations of visual data, research targeting an efficient compression of 3D point clouds received more and more attention \revision{\cite{Liu19}}. Examples for standardization activities are geometry-based point cloud compression (G-PCC) \cite{MPEG_GPCC},\revision{\cite{Liu21}} and video-based point cloud compression (V-PCC) \cite{Graziosi20, MPEG_VPCC}. As the storage space and data rates for the transmission of point clouds are extremely high, these two activities were launched to develop algorithms for the efficient compression of raw point cloud data. 

For G-PCC, the general idea is to compress the data in the 3D voxel domain. As a result, the geometry of the voxels is coded losslessly. In contrast,  for V-PCC, the point cloud is projected onto various 2D image planes such that traditional lossy video compression can be used for the compression of the 2D projections. As a consequence, the geometry as well as the texture is subject to distortions. 

\begin{figure}[t]
\centering
\includegraphics[width=.49\textwidth]{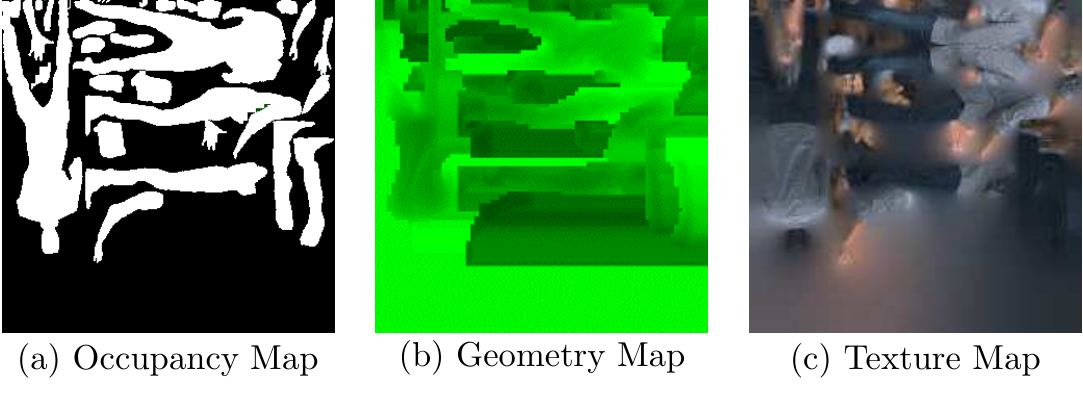}
\vspace{-0.7cm}
\caption{2D images created from a 3D point cloud. The occupancy map (a) is compressed losslessly, the geometry map (b) and the texture map (c) are compressed in a lossy way. The algorithm proposed in this work targets the latter two maps. } 
\label{fig:PCCimages}
\end{figure}

During the projection of the point cloud to 2D images, so-called patches are generated by orthonormal projection of the original point cloud, which have an arbitrary shape \cite{Li20}. Then, all information is packed into three 2D images: First, a so-called occupancy map as shown in Fig.~\ref{fig:PCCimages}~(a) indicates which pixels of the images contain relevant information. In this representation, white corresponds to occupied pixels, i.e., relevant information, and black to unoccupied pixels. Second, the geometry map in Fig.~\ref{fig:PCCimages}~(b) indicates the distance of the corresponding pixel from the virtual camera, such that the 3D voxel coordinates can be reconstructed from the x-position, the y-position, and the brightness of the current pixel representing the depth. Third, the texture map as illustrated in Fig.~\ref{fig:PCCimages}~(c) assigns color information to each voxel. 

The proposed method is based on occupancy-map-based rate distortion optimization \cite{Li20} and inactive region coding for $360^\circ$ videos \cite{Herglotz19c}. In addition to the aforementioned work, a novel algorithm is introduced for determining rate-distortion-optimized residual coefficients for intra and inter coded frames. The method is performed on all blocks containing both occupied and unoccupied pixels (mixed blocks). For blocks consisting of purely occupied pixels, the classic RDO approach is performed, and for purely unoccupied blocks, the residual error is set to zero, as proposed in \cite{Herglotz19c}. For mixed blocks, the goal of our proposed method is to determine optimal transformed, residual coefficients, which minimize the distortion in the occupied region while neglecting the distortion in the unoccupied region, and simultaneously minimizing the rate. In this respect, the problem is interpreted as a signal extrapolation task, where the relevant signal is extrapolated to the unoccupied region. 

The novel contributions of this paper are as follows: 
\begin{itemize}
\item Proposal of a dedicated selective extrapolation algorithm for efficient compression of unoccupied regions. 
\item Introduction of a rate-distortion based selection criterion to enhance the performance of classical selective extrapolation using rate estimation. 
\item  Simplification of the iterative procedure by exploiting sparsity. 
\end{itemize}

This paper is structured as follows: First, Section~\ref{sec:lit} reviews the current state of the art in V-PCC and on compression of unoccupied pixels as well as frequency selective extrapolation. 
Afterwards, Section~\ref{sec:ose} discusses the state-of-the-art approaches for the determination and compression of residual signals and proposes an optimal selective extrapolation algorithm including the rate-distortion optimization procedure. 
Afterwards, Section~\ref{sec:eval} introduces the test data set and performs a detailed evaluation of the coding gains. Finally, Section~\ref{sec:concl} concludes this paper.

\section{Literature Review}
\label{sec:lit}

In this section, we discuss the state-of-the-art compression techniques used in video-based point cloud compression and the handling of unoccupied pixels. 
 Finally, the basics for our proposed rate-constrained optimal selective extrapolation (ROSE) are introduced. 

\subsection{Video-based Point Cloud Compression}
\begin{figure}[t]
\centering
\includegraphics[width=.45\textwidth]{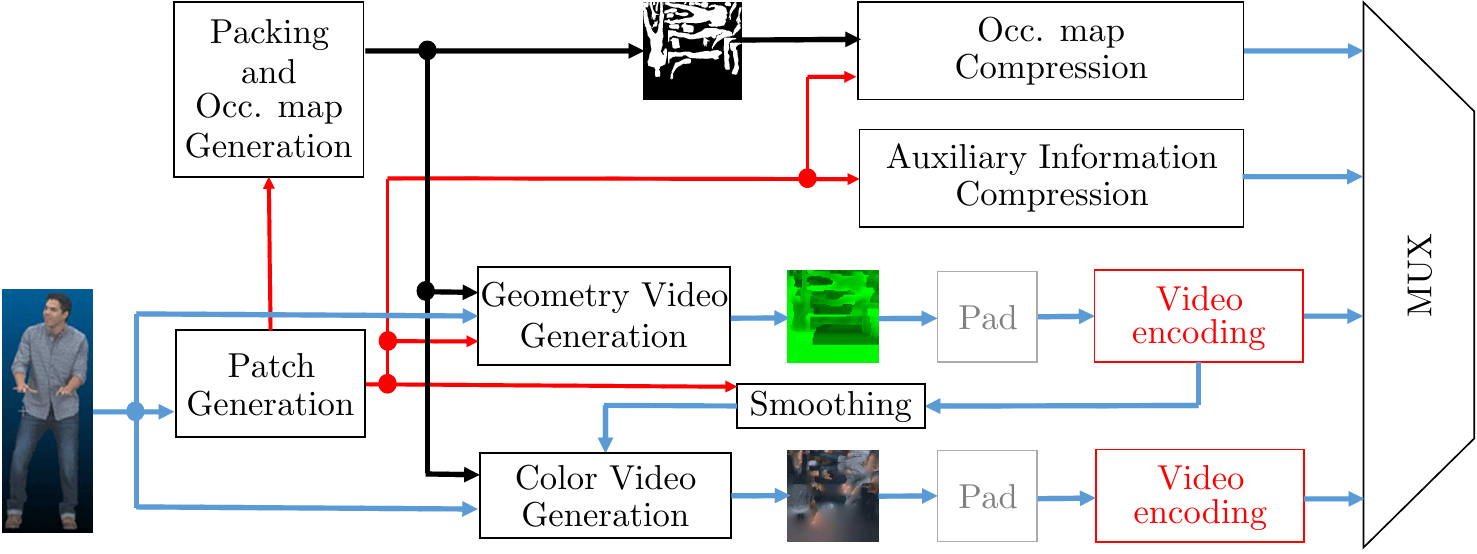}
\vspace{-.3cm} 
\caption{\revision{Generation and compression of 2D-videos in point cloud compression based on \cite{MPEG_VPCC}. 2D patches (red arrows) are generated from an input point cloud, which are used to generate an occupancy map (black arrows), a geometry video, and a texture video. Then, the two videos are padded and encoded with standard video compression. The output bit streams are multiplexed with the compressed occupancy map and the compressed auxiliary patch information. 
Our proposed algorithm is inserted in video encoding for the geometry video and the color video (red boxes).}   } 
\label{fig:VPCC-schematic}
\end{figure}

In video-based point cloud compression (V-PCC) \cite{Graziosi20,MPEG_VPCC}, a given point cloud with texture information is decomposed into multiple components, which are then compressed with standard video compression techniques. A general overview of the components relevant for video compression is depicted in Fig.~\ref{fig:VPCC-schematic}. 

The main idea is that multiple two-dimensional patches from the point cloud are generated by orthographic projection onto multiple two-dimensional planes. The positions of the corresponding virtual cameras are chosen in such a way that the complete point cloud is covered. Then, the resulting patches are packed onto two-dimensional images, which can be further compressed by standard image and video compression techniques. Additional data such as the position of the cameras is signaled as side information. 

This projection process creates three videos: The first one is a so-called occupancy map, which indicates with a binary mask whether a pixel in the video holds relevant information or not. The occupancy map is usually coded with a lossless encoder. 

The second video is the so-called geometry map. For each occupied pixel, this map indicates the depth of the corresponding point in the point cloud. This depth is saved in the Y-luminance channel and compressed with a standard video encoder. 

The third video is the texture map which holds information about texture properties such as brightness, color, or other properties such as the reflectance or tactile properties of voxels. In this paper, we focus on RGB color information, which is represented in the widely used YCbCr 4:2:0 color format.


For video encoding, the state-of-the-art approach is to use HEVC \cite{Graziosi20}. However, it is pointed out that any image or video compression algorithm could be used. For example, a learning-based method for lossless compression was proposed in \cite{Nguyen21} and \cite{Guarda19}. 

Other work focused on optimizing the framework for more efficient encoding. For example, it was proposed to simplify the grid-based refining segmentation to reduce processing times \cite{Kim21}. It was found that the compression performance in terms of rate-distortion costs did not suffer. Another approach is to exploit temporal correlations between the near and the far layer, which leads to more than $40\%$ time savings at marginal increases in bitrate \cite{Xiong21}. Finally, an artifact removal framework for post-processing of reconstructed point clouds was presented and it was shown that the reconstruction quality could be increased significantly without changing the bitrate \cite{Akhtar21}. 

\revision{Finally, several works target the rate-distortion (RD) process in V-PCC. For example, \cite{Liu21b} propose to employ a perceptual quality model to improve rate control. Furthermore, it is proposed to use %
%
a distortion and a rate model jointly to adaptively change the QP for geometry and color encoding to accurately reach a target bitrate. To achieve this, in \cite{Yuan21} it was proposed to use a method based on differential evolution and in \cite{Liu20}, an interior point method was chosen. 
At small losses in compression efficiency, lower errors with respect to the target bitrate are achieved. In a similar direction, Yuan et al. proposed to improve the RD performance using differential evolution by removing the constraint of a fixed QP and adaptively changing the QP on frame level \cite{Yuan21b}. 
}


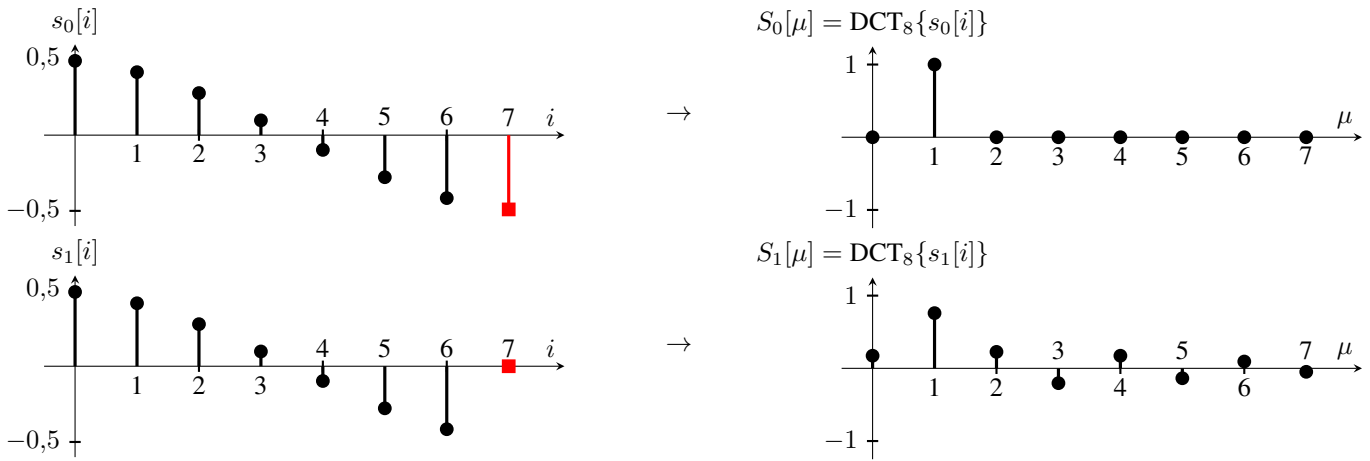
\begin{figure*}[h!t]
	\begin{minipage}{0.45\textwidth}
		\begin{tikzpicture}
		\begin{axis}[%
		SISYstyle,
		width=8.5cm,
		height=4cm,
		axis x line=middle,
		xmin=-0.5, xmax=7.9,
		ymin=-0.6, ymax=0.6,
		xlabel={$i$},
		ylabel={$s_0[i]$},
		xticklabels=\empty,
		/pgf/number format/.cd,
		use comma
		]
		\addplot+[ycomb,black,very thick,mark options={fill=black}] plot coordinates {(0,0.4904) (1, 0.4157) (2,0.2778) (3,0.0975) (4,-0.0975) (5,-0.2778) (6,-0.4157)};
		\addplot+[ycomb,red,very thick,mark options={fill=red}] plot coordinates {(7,-0.4904)};
		\node[below] at (axis cs:1,0) {1}; 
		\node[below] at (axis cs:2,0) {2}; 
		\node[below] at (axis cs:3,0) {3}; 
		\node[above] at (axis cs:4,0) {4}; 
		\node[above] at (axis cs:5,0) {5}; 
		\node[above] at (axis cs:6,0) {6}; 
		\node[above] at (axis cs:7,0) {7}; 
		\end{axis}
		\end{tikzpicture}
	\end{minipage}
	\hfill
	\begin{minipage}{0.05\textwidth}
		\centering
		$\rightarrow$
	\end{minipage}
	\hfill
	\begin{minipage}{0.45\textwidth}
		\begin{tikzpicture}
		\begin{axis}[%
		SISYstyle,
		width=8.5cm,
		height=4cm,
		axis x line=middle,
		xmin=-0.5, xmax=7.9,
		ymin=-1.25, ymax=1.25,
		xlabel={$\mu$},
		ylabel={$S_0[\mu] = \text{DCT}_8 \{ s_0[i] \} $},
		xticklabels=\empty,
		/pgf/number format/.cd,
		use comma
		]
		\addplot+[ycomb,black,very thick,mark options={fill=black}] plot coordinates 
		{(0,0) (1, 1) (2,0) (3,0) (4,0) (5,0) (6,0) (7,0)};
		\node[below] at (axis cs:1,0) {1}; 
		\node[below] at (axis cs:2,0) {2}; 
		\node[below] at (axis cs:3,0) {3}; 
		\node[below] at (axis cs:4,0) {4}; 
		\node[below] at (axis cs:5,0) {5}; 
		\node[below] at (axis cs:6,0) {6}; 
		\node[below] at (axis cs:7,0) {7}; 
		\end{axis}
		\end{tikzpicture}
	\end{minipage}
	\begin{minipage}{0.45\textwidth}
		\begin{tikzpicture}
		\begin{axis}[%
		SISYstyle,
		width=8.5cm,
		height=4cm,
		axis x line=middle,
		xmin=-0.5, xmax=7.9,
		ymin=-0.6, ymax=0.6,
		xlabel={$i$},
		ylabel={$s_1[i]$},
		xticklabels=\empty,
		/pgf/number format/.cd,
		use comma
		]
		\addplot+[ycomb,black,very thick,mark options={fill=black}] plot coordinates {(0,0.4904) (1, 0.4157) (2,0.2778) (3,0.0975) (4,-0.0975) (5,-0.2778) (6,-0.4157)};
		\addplot+[ycomb,red,very thick,mark options={fill=red}] plot coordinates {(7,0)};
		\node[below] at (axis cs:1,0) {1}; 
		\node[below] at (axis cs:2,0) {2}; 
		\node[below] at (axis cs:3,0) {3}; 
		\node[above] at (axis cs:4,0) {4}; 
		\node[above] at (axis cs:5,0) {5}; 
		\node[above] at (axis cs:6,0) {6}; 
		\node[above] at (axis cs:7,0) {7}; 
		\end{axis}
		\end{tikzpicture}
	\end{minipage}
	\hfill
	\begin{minipage}{0.05\textwidth}
		\centering
		$\rightarrow$
	\end{minipage}
	\hfill
	\begin{minipage}{0.45\textwidth}
		\begin{tikzpicture}
		\begin{axis}[%
		SISYstyle,
		width=8.5cm,
		height=4cm,
		axis x line=middle,
		xmin=-0.5, xmax=7.9,
		ymin=-1.25, ymax=1.25,
		xlabel={$\mu$},
		ylabel={$S_1[\mu] = \text{DCT}_8 \{ s_1[i] \} $},
		xticklabels=\empty,
		/pgf/number format/.cd,
		use comma
		]
		\addplot+[ycomb,black,very thick,mark options={fill=black}] plot coordinates 
		{(0,0.1734) (1, 0.7595) (2,0.2265) (3,-0.2039) (4,0.1734) (5,-0.1362) (6,0.0938) 
			(7,-0.0478)};
		\node[below] at (axis cs:1,0) {1}; 
		\node[below] at (axis cs:2,0) {2}; 
		\node[above] at (axis cs:3,0) {3}; 
		\node[below] at (axis cs:4,0) {4}; 
		\node[above] at (axis cs:5,0) {5}; 
		\node[below] at (axis cs:6,0) {6}; 
		\node[above] at (axis cs:7,0) {7}; 
		\end{axis}
		\end{tikzpicture}
	\end{minipage}
	\caption{The impact of unoccupied regions shown in spatial and transform domain for a 1-D example. Black dots are relevant samples, while red squares are assumed to be irrelevant, i.e., unoccupied. The signals $s_0[i]$ and $s_1[i]$ are transformed to the {DCT-II} domain for obtaining $S_0[\mu]$ and $S_1[\mu]$, which are quantized and coded afterwards. Already a single unoccupied pixel can lead to a significant energy decompaction of transform coefficients (approx. $21$\,\% for $s_1[i]$). Hence, a large distortion must be accepted after quantization for $s_1[i]$, while $s_0[i]$ would also lead to an optimal result.}
	\label{img:energy-decompaction}
\end{figure*}

\subsection{Occupancy-Map-Based Compression}
From Fig.~\ref{fig:VPCC-schematic}, one can infer that the generated 2D videos inhibit large areas which do not contain relevant information to reconstruct the original point clouds. However, during compression, such regions contribute significantly to the resulting bitrate, which was pointed out in various works \cite{Graziosi20, Li20, Herglotz19c}. To mitigate the impact of these regions on the overall bitrates, different techniques were used and proposed. 

The standard proposal from V-PCC uses a pixel padding technique, which, depending on the block position, performs dilation or copying of pixels to obtain smooth transitions \cite{Graziosi20}. For other input data, different padding techniques were discussed and evaluated for different projection formats in $360^\circ$-video coding \cite{JVET_K1004}. Finally, different approaches for object-based texture coding were discussed and evaluated in \cite{Kaup99}, where also different transforms were investigated. 

Another idea is to extrapolate the useful signal information into unoccupied regions, where an approach for non-rectangular signals was proposed in \cite{Das20}. The method uses an orthogonal matching pursuit algorithm to extend the data to rectangular blocks. 
As another idea, it was proposed in \cite{Herglotz19c} to reduce the energy of the residual signal by setting residual pixel values located in unoccupied regions to zero, before they are transformed by a discrete cosine transform (DCT). Significant bitrate savings of up to $10\%$ were reported for $360^\circ$-videos \cite{Herglotz19c}. 

For point clouds, it was proposed to change the encoder's functionality in such a way that during the calculation of rate-distortion costs, the distortion in unoccupied regions is neglected when calculating the sum of squared errors (SSE) \cite{Li20}. This principle was applied for calculating accurate RD costs in intra prediction, inter prediction, and in the selection of the sample-adaptive offset (SAO) filter coefficients. Rate savings of up to $10\%$ were reported.

\subsection{Selective Extrapolation with Arbitrary Basis Functions}
\label{sec:SE}
Coding of irrelevant or unoccupied regions can be intuitively formulated as a sparse signal extrapolation task. In signal extrapolation, greedy sparse algorithms are frequently used, especially, when computation time plays a crucial role \cite{olshausen97}. A versatile signal extrapolation method is Frequency Selective Extrapolation (FSE) introduced in \cite{kaup2005}. In contrast to other signal extrapolation algorithms like \cite{dahl2009} or \cite{starck2010}, FSE is in continuous development and it has been refined by several extensions to improve extrapolation quality \cite{koloda2014} and to reduce computational complexity \cite{genser2018}. Moreover, it was shown in \cite{genser2017} that fast implementations can be formulated in a transform domain that are real-time capable. Recently, also the gap between compressed sensing and FSE has been closed \cite{grosche2020}, where it was demonstrated that the FSE framework is able to outperform state-of-the-art algorithms such as Approximate Message Passing \cite{donoho2009}, Orthogonal Matching Pursuit \cite{tropp2007}, and its successors. While most studies of FSE were conducted for Fourier-based models, it was shown in \cite{seiler2011} that the algorithm is also well suited for using arbitrary sets of basis functions. As recent video codecs such as High Efficiency Video Coding (HEVC) or Versatile Video Coding (VVC) are known to utilize plenty of transforms, the formulation of a basis function-independent algorithm is also of high interest in the coding of unoccupied regions. 


\section{Rate-Constrained Optimal Selective Extrapolation}
\label{sec:ose}

\revision{In the following, we motivate the general idea of our proposed approach. 
When coding unoccupied regions, the unnecessary pixels must also be assigned to numeric values even if these samples are not required in the latter processing. However, their choice has a significant impact on the coding efficiency. Assuming the simplified case of a one-dimensional signal, this effect is visualized in Fig.~\ref{img:energy-decompaction}, which shows a comparison of two exemplary signals in the spatial as well as in the transform domain. Two signals $s_0[i]$ and $s_1[i]$ are transformed to the {DCT-II} domain which results in the transformed signals $S_0[\mu]$ and $S_1[\mu]$. The variables $i$ and $\mu$ denote the spatial and frequency indices, respectively. From Fig.~\ref{img:energy-decompaction}, one can infer that already a single unoccupied pixel can lead to a widespread distribution of transform coefficients. This leads to a suboptimal energy compaction, and consequently to a disadvantageous distribution of residual coefficients. }
\par
\revision{A widespread distribution of residual coefficients impairs the compression performance 
 \cite{Sole12}. The reason is the common order of signaling, which follows an inverse zigzag scan beginning from the last non-zero coefficient in a block of residual coefficients. If the last nonzero coefficient is located close to the DC coefficient, only few coefficients need to be encoded. If the last nonzero coefficient has a high frequency, the opposite is true. In the given example, signal $S_0[\mu]$ can be encoded more efficiently because only one coefficient has to be transmitted (position $\mu=1$). For signal $S_1[\mu]$, depending on the quantization, up to eight coefficients must be transmitted. 
}
\par
In the following, we derive a general problem formulation to code unoccupied regions in such a way that its coding does not notably increase the overall bit stream size. This is achieved by replacing unoccupied samples so that they hold a compact energy model in the selected transform domain. In order to motivate our proposed Rate-constrained Optimal Selective Extrapolation (ROSE) approach, the impact of unoccupied regions is first analyzed in spatial as well as in transform domain. After discussing the fundamental problem, the novel ROSE algorithm is introduced as a possible solution, where a rate-distortion criterion is included as the optimization function. 
Finally, the algorithmic design is completed by deriving a fast ROSE algorithm, which achieves a significantly lower computational complexity at the same coding performance of conventional ROSE.

\subsection{Residual Signal Compression}
\begin{figure}[t]
\centering
\includegraphics[width=.49\textwidth]{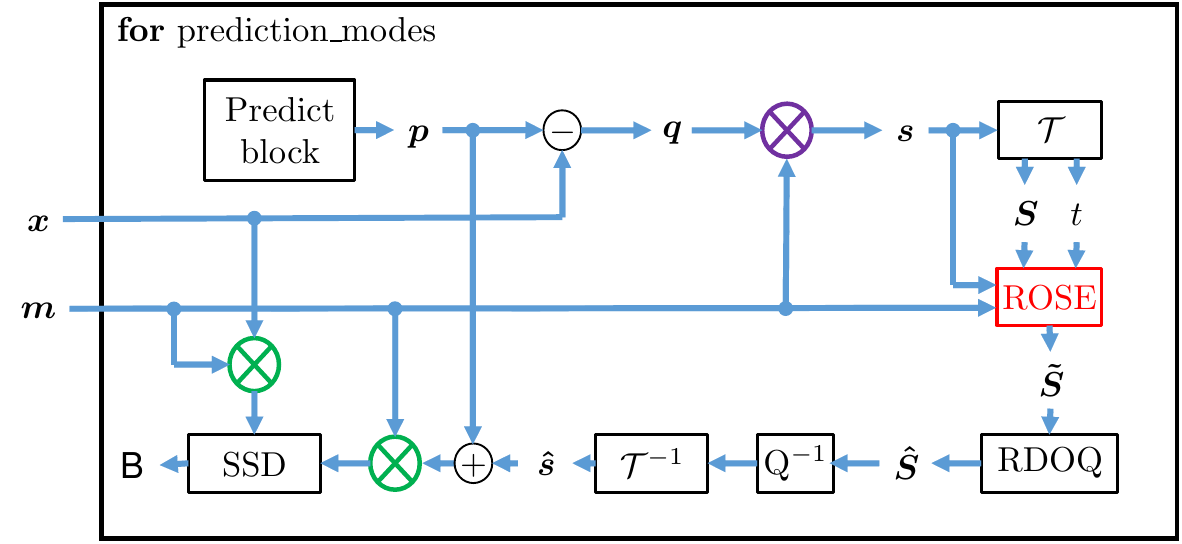} 
\caption{Signal flow of the residual signal $\boldsymbol q$ in rate-distortion optimization for different prediction modes. The input image data is $\boldsymbol x$ and the mask is denoted as $\boldsymbol m$. The black blocks represent the original encoding flow, the colored blocks and operators correspond to additional functionality for optimized unoccupied region coding. The position of the proposed ROSE approach is shown in red.  }
\label{fig:resSigFlow}
\end{figure} 

Fig.~\ref{fig:resSigFlow} shows the RD optimization of the HM encoder implementation of HEVC \cite{HM}, which determines the rate-distortion costs of the residual coefficients given a certain prediction signal. 
 Note that this process is valid for any block size ($4\times 4$ up to $32\times 32$), where the decision for the final block size is performed based on the minimum RD-costs of all potential block partitionings. 

 First, the current block is predicted by any of the available intra and inter prediction modes, which results in the predicted pixel values $\boldsymbol{p}$. This prediction is subtracted from the original signal $\boldsymbol{x}$, which contains the uncompressed input pixel values to obtain the residual error signal $\boldsymbol{q} = \boldsymbol{p} - \boldsymbol{x}$. In standard encoding, this residual signal $\boldsymbol{q}$ is equal to the input signal of the transform $\boldsymbol{s}$. For optimized compression results, it was proposed in \cite{Herglotz19c} to multiply $\boldsymbol{q}$ with the occupancy map $\boldsymbol{m}$ before transformation in order to reduce the bitrate (purple multiplicator), which will be called residual masking (RM) in the following. 

\revision{After transformation with a transform type $t$ of the set of available transforms $\mathcal{T}$}, we obtain the masked residual signal in the frequency domain $\boldsymbol{S}$, which will be further refined by our proposed ROSE approach to obtain the signal $\boldsymbol{\tilde S}$.  Afterwards, $\boldsymbol{\tilde S}$ is quantized by rate-distortion optimized quantization (RDOQ) resulting in the quantized coefficients $\boldsymbol{\hat S}$, which are fed into the context-adaptive binary arithmetic coding (CABAC) to determine the bitrate (not shown). 

Then, the signal is scaled (inverse quantization) and inversely transformed ($\mathcal{T}^{-1}$) to obtain the residual error in spatial domain after quantization $\boldsymbol{\hat s}$, which is added to the predicted signal $\boldsymbol{p}$. In standard encoding, the resulting signal is used to calculate the distortion in terms of the sum of squared differences (SSD), which is then used to calculate the RD-costs. In \cite{Li20} and \cite{Herglotz19c}, it was proposed to calculate this distortion only for the occupied pixels, which is achieved by multiplying both the residual signal $\boldsymbol{\hat s}$ and the original signal $\boldsymbol{x}$ with the unoccupied pixel mask $\boldsymbol{m}$ (green multipliers). In the following, this approach will be called occupied distortion (OD) calculation.



\revision{In the proposed ROSE approach, we optimize the residual coefficients for each tested coding mode on each tested block size when both occupied and unoccupied pixels are present in the current block. As such, we perform an exhaustive search for the best compression performance. }
\par
Apparently, \revision{as visualized in Fig.~\ref{img:energy-decompaction},} the coding of unoccupied regions can be interpreted as a signal extrapolation problem with the constraints of generating a highly sparse model in the transform domain, while favoring transform coefficients that yield low coding costs. This leads us to the design of the ROSE approach as discussed in the following.

\subsection{Basis Function Derivation}
\label{sec:OSE-bf}
In order to achieve an efficient compression of the spatial signal, coding schemes such as HEVC apply different content-adaptive transforms to each Transform Unit (TU). In the following, ROSE will be derived for HEVC, however, it is applicable to every transform based video coder. The set
\begin{align}
	\mathcal{T} = \{ \text{DST}_4, \text{DCT}_4, \text{DCT}_8, \text{DCT}_{16}, \text{DCT}_{32} \}
	\label{eq:trafos}
\end{align}
stores all possible transform types $t \in \mathcal{T}$ of the HEVC codec.
\par
The previously introduced effect of energy decompaction varies in strength and behavior depending on the selected transform. In line with this, a general strategy is required to code unoccupied regions independent of the applied transform type. To this end, ROSE models a given signal as a linear combination of arbitrary basis functions and does not demand for further knowledge of the applied coding scheme. It achieves an optimal approximation of the relevant pixels as long as the applied transform is known beforehand. The basis functions required by ROSE are obtained by inverse transform of the applied type $t \in \mathcal{T}$ for each spectral coefficient. For a separable two-dimensional transform as in HEVC, we obtain the transformed signal
\begin{align}
	\boldsymbol{S}_t = \boldsymbol{T}_t^\intercal \boldsymbol{s}_t \boldsymbol{T}_t 
\end{align}
with respect to the a priori known transform matrices $\boldsymbol{T}_t$ of type $t$ in the set of all possible HEVC transforms $\mathcal{T}$. Accordingly, the inverse transform can be depicted as
\begin{align}
	\boldsymbol{s}_t = (\boldsymbol{T}_t^\intercal)^{-1} \boldsymbol{S}_t (\boldsymbol{T}_t)^{-1} \text{ .}
\end{align}
For orthogonal transforms, it yields
\begin{align}
	\boldsymbol{T}_t^{-1} = \boldsymbol{T}_t^\intercal \text{ ,}
\end{align}
which simplifies the calculation of the inverse transform such that the backward transforms result in
\begin{align}
	\boldsymbol{s}_t = \boldsymbol{T}_t \boldsymbol{S}_t \boldsymbol{T}_t^\intercal \text{ .}
\end{align}
In order to derive the basis functions in the spatial domain, each coefficient of type $t$ must be inversely transformed. For all two-dimensional coefficient indices $\mu_1$ and $\mu_2$, we obtain the basis functions in transform domain
\begin{align}
	\boldsymbol{\mathit{\Phi}}_{t,k_1k_2} =
		\begin{cases}
			1, & \text{ for } k_1 = \mu_1 \land k_2=\mu_2\\
			0, & \text{ else }
		\end{cases} \text{ , \quad} \forall k_1, k_2 \text{ ,}
\end{align}
where the indices $k_1$ and $k_2$ indicate the 2D position of the non-zero value. 
By applying the inverse transform of type $t$, the according basis function in spatial domain results in
\begin{align}
	\boldsymbol{\varphi}_{t,k_1,k_2} = \boldsymbol{T}_t \boldsymbol{\mathit{\Phi}}_{t,k_1,k_2} \boldsymbol{T}_t^\intercal \text{ , \quad} \forall k_1, k_2 \text{ .}
\end{align}
The two-dimensional functions $\boldsymbol{\varphi}_{t,k_1,k_2}$ are then rewritten as one-dimensional vectors $\boldsymbol{\varphi}_{t,k}$ with indices
\begin{align}
	\mu = \mu_1+B\mu_2 \text{\quad and \quad} k = k_1+Bk_2
	\label{eq:2Dto1D}
\end{align}
and $B$ denoting the width of the applied transform, e.g., $B_{\text{DCT}_4} = 4$, and $B_{\text{DCT}_{16}} = 16$. 
By including all basis functions into the model generation of ROSE, the complete dictionary of basis functions of transform type $t$ is denoted as
\begin{align}
	\mathcal{D}_t = \bigcup_{k \, \in \, \mathcal{K}_t} \boldsymbol{\varphi}_{t,k} \text{ .}
	\label{eq:dictionary}
\end{align}
In above equation, set $\mathcal{K}_t$ holds all possible indices $k$ which depend on the transform type $t$. For example, the DCT$_4$ transform leads to $|\mathcal{K_{\text{DCT$_{4}$}}}|=4\times 4 = 16$ basis functions, while a type DCT$_{32}$ results in $|\mathcal{K_{\text{DCT$_{32}$}}}| = 32\times 32 = 1024$ basis functions.
\par
The derivation of the basis function sets must be conducted only once for all coded TUs. In practical applications, it is highly recommendable to store the sets in precalculated look-up tables so that computational complexity is not increased compared to conventional coding.

\subsection{Rate-Constrained Optimal Selective Extrapolation}

After transformation, 
the sparse signal $\boldsymbol{{S}}$ is obtained in the encoder (see Fig.~\ref{fig:resSigFlow}). When introducing unoccupied pixels, $\boldsymbol{{S}}$ may suffer from energy decompaction. Thus, the goal must be to estimate a model which does not take the unoccupied pixels into account, while maintaining the relevant samples and being as sparse as possible to keep the coding rate costs low. As the unoccupied region is defined in the spatial domain, the problem can be modeled as
\begin{align}
	\boldsymbol{g}_t = \sum_{k \in \mathcal{\hat{K}}_t} \hat{c}_k \boldsymbol{\varphi}_{t,k}
\end{align}
with estimated coefficients $\hat{c}_k$ and a set of selected basis functions $\mathcal{\hat{K}}_t 
\subset \mathcal{K}_t$. The problem definition, which we solve in a rate-distortion optimal sense, can be divided into two sub-tasks: First, an optimal coefficient must be selected. Second, the 
strength of the chosen coefficient has to be estimated. As both sub-tasks are a joint optimization problem, ROSE 
is formulated as an iterative approximation algorithm. For keeping the algorithmic formulation short and
precise, the transform type $t$ is omitted in the following since the model generation is always 
conducted for one specific transform type $t$ at a time. When discussing, e.g., $\boldsymbol{g}$, implicitly $\boldsymbol{g}_t$ is intended. First, the initial model
\begin{align}
	\boldsymbol{g}_{0,i} = 0 \text{ , } \forall i
\end{align}
as well as the residual
\begin{align}
	\boldsymbol{r}_{0,i} = \boldsymbol{s}_{i} \text{ , } \forall i
\end{align}
are initialized for iteration $\nu = 0 \text{ with } \nu \in \{0, ..., N-1\}$. Similar to eq. \eqref{eq:2Dto1D}, $i$ is the one-dimensional spatial index that is derived from the two-dimensional block position $(i_1,i_2)$. The number of iterations of $N$ is chosen by counting the number of non-zero coefficients after quantization in the encoder, which leads to the $l_0$ norm
\begin{align}
	N = | \boldsymbol{{S}} |_0
	\label{eq:density}
\end{align}
and can be interpreted as density of $\boldsymbol{{S}}$. Consequently, the proposed ROSE approach always uses at maximum as many coefficients as the conventional encoder, but it chooses the coefficients such that only the relevant regions are optimally approximated, which leads to a better signal model at approximately the same rate. For each iteration $\nu$, the preliminary projection coefficients
\begin{align}
	\boldsymbol{\check{c}}_{\nu,k} = \dfrac{\sum\limits_{i} \boldsymbol{r}_{\nu,i} 
	\boldsymbol{\varphi}_{k,i} 
	\boldsymbol{m}_i}{\sum\limits_{i} \boldsymbol{\varphi}_{k,i}^2 \boldsymbol{m}_i} \text{ , } 
	\forall k
	\label{eq:coefficient-estimation}
\end{align}
are estimated. In above equation, vector $\boldsymbol{m}$ denotes the a priori defined occupancy map, which 
takes the value $1$ for a relevant pixel and $0$ for an unoccupied sample. Given the projection coefficients
$\boldsymbol{\check{c}}_{\nu,k}$, the model error
\begin{align}
	 \boldsymbol{e}_{\nu,k} = \sum\limits_{i} | \boldsymbol{r}_{\nu-1,i} - 
	 \boldsymbol{\check{c}}_{\nu,k} 
	 \boldsymbol{\varphi}_{k,i} |^2 \boldsymbol{m}_i \text{ , } \forall k \label{eq:modelError}
\end{align}
is calculated to determine the quality of the estimated coefficient for each basis function and the current model.
In line with this, the best coefficient index $\hat{k}_{\nu}$ for iteration $\nu$ results in
\begin{align}
	\hat{k}_{\nu} = \argmin\limits_{k} \left\{ \boldsymbol{e}_{\nu} + \lambda \cdot b(\boldsymbol{\check c})\right\}
	\label{eq:coeff-sel}
\end{align}
by minimizing the joint coding costs of the above error function and the product of a Lagrange multiplier $\lambda$ and the bitrate $b(\boldsymbol{\check c})$, which will both be discussed in the subsequent section. After selecting the best coefficient index, the model is 
optimized for all already selected 
coefficients in set $\mathcal{\hat{K}}_{\nu}$ by deriving the set of linear equations
\begin{align}
	\boldsymbol{M}
	\begin{bmatrix}
	\boldsymbol{\varphi}_{\hat{k}_{0}}\\ \vdots\\ \boldsymbol{\varphi}_{\hat{k}_{\nu}}
	\end{bmatrix}^\intercal
	\begin{pmatrix}
		\boldsymbol{\hat{c}}_{0}\\
		\vdots\\
		\boldsymbol{\hat{c}}_{\nu}
	\end{pmatrix}
	=
	\boldsymbol{M} \boldsymbol{s} \text{ ,}
	\label{eq:global-LSE}
\end{align}
with $\boldsymbol{M} = \mathrm{diag}(\boldsymbol{m})$ being the according diagonal matrix of the 
unoccupied region mask $\boldsymbol{m}$. For an orthogonal set of basis functions, the estimation of coefficients can be independently conducted from each other. As the multiplication by a mask leads to an orthogonal deficiency, eq. (\ref{eq:global-LSE}) allows for a joint problem formulation of all already selected basis functions.
In contrast to the preliminary projection coefficients $\boldsymbol{\check{c}}$, the estimated coefficients 
$\boldsymbol{\hat{c}}$ shall be determined to minimize the error between the selected basis 
functions 
and the known pixels on the relevant mask positions in the least-square sense for all already 
selected 
basis 
functions at a time, so that
\begin{align}
	\begin{bmatrix}
	\boldsymbol{\varphi}_{\hat{k}_{0}}\\ \vdots\\ \boldsymbol{\varphi}_{\hat{k}_{\nu}}
	\end{bmatrix}
	\boldsymbol{M}
	\begin{bmatrix}
	\boldsymbol{\varphi}_{\hat{k}_{0}}\\ \vdots\\ \boldsymbol{\varphi}_{\hat{k}_{\nu}}
	\end{bmatrix}^\intercal
	\begin{pmatrix}
	\boldsymbol{\hat{c}}_{0}\\
	\vdots\\
	\boldsymbol{\hat{c}}_{\nu}
	\end{pmatrix}
	=
	\begin{bmatrix}
	\boldsymbol{\varphi}_{\hat{k}_{0}}\\ \vdots\\ \boldsymbol{\varphi}_{\hat{k}_{\nu}}
	\end{bmatrix}
	\boldsymbol{M} \boldsymbol{s}
\end{align}
must be solved. By taking the Moore-Penrose inverse into account, the estimated coefficients 
result 
in
\begin{align}
	\begin{pmatrix}
		\boldsymbol{\hat{c}}_{0}\\
		\vdots\\
		\boldsymbol{\hat{c}}_{\nu}
	\end{pmatrix}
	=
	\left(
	\begin{bmatrix}
	\boldsymbol{\varphi}_{\hat{k}_{0}}\\ \vdots\\ \boldsymbol{\varphi}_{\hat{k}_{\nu}}
	\end{bmatrix}
	\boldsymbol{M}
	\begin{bmatrix}
	\boldsymbol{\varphi}_{\hat{k}_{0}}\\ \vdots\\ \boldsymbol{\varphi}_{\hat{k}_{\nu}}
	\end{bmatrix}^\intercal
	\right)^{-1}
	\begin{bmatrix}
	\boldsymbol{\varphi}_{\hat{k}_{0}}\\ \vdots\\ \boldsymbol{\varphi}_{\hat{k}_{\nu}}
	\end{bmatrix}
	\boldsymbol{M} \boldsymbol{s}
 \text{ .}
 \label{eq:coeffUpdate}
\end{align}
Consequently, the model at iteration $\nu$ is given as
\begin{align}
	\boldsymbol{g}_{\nu} =
	\begin{bmatrix}
		\boldsymbol{\varphi}_{\hat{k}_{0}}\\
		\vdots\\
		\boldsymbol{\varphi}_{\hat{k}_{\nu}}
	\end{bmatrix}^\intercal
	\begin{pmatrix}
		\boldsymbol{\hat{c}}_{0}\\
		\vdots\\
		\boldsymbol{\hat{c}}_{\nu}
	\end{pmatrix} \text{ .}
	 \label{eq:modelUpdate}
\end{align}
In line with this, the residual
\begin{align}
	\boldsymbol{r}_{\nu+1} = \boldsymbol{s} - \boldsymbol{g}_{\nu}
\end{align}
is updated and the subsequent iteration can be carried out. The algorithm is conducted 
$N$ times, until the sparse model is estimated. However, $N$ is typically very small in practical 
applications due to the applied quantization in the video coder.
\par
Finally, the transform signal is depicted as
\begin{align}
	\boldsymbol{\tilde{S}} =
		\begin{cases}
			\boldsymbol{\hat{c}}_{k}, & k \in \mathcal{\hat{K}}\\
			0, & \text{else }
		\end{cases}
\end{align}
and replaces the conventional transform signal $\boldsymbol{S}$ of the video encoder.

\begin{figure}[t]
	\centering
	\hspace*{-1em}
	\begin{tikzpicture}
	\small
	\tikzstyle{box} = [draw]
	
	\node[] (in1) at (2,2) {\begin{tabular}{c} 
	Transformed signal $\boldsymbol{{S}}$\\
		Signal $\boldsymbol{s}$\\
		Mask $\boldsymbol{m}$
		\end{tabular}};
	\node[] (in2) at (6,2) {\begin{tabular}{c}Transform type $t$\end{tabular}};
	\node[box,dashed,minimum size=8cm,minimum width=8cm,label={[xshift=3cm,yshift=-0.35cm,above]\scriptsize\colorbox{white}{Proposed ROSE}}] (Box) at (4,-3) {};
	\node[box, minimum size=1cm,minimum width=2cm] (cca) at (2,0) {\begin{tabular}{c}Conventional coding\\density analysis (\ref{eq:density})\end{tabular}};
	\node[box, minimum size=1cm,minimum width=2cm] (bfd) at (6,0) {\begin{tabular}{c}Basis function\\derivation (\ref{eq:dictionary})\end{tabular}};
	\node[draw,diamond, aspect=1.8] (fin) at (2,-2) {\begin{tabular}{c}Finished?\\$\nu < N$\end{tabular}};
	\node[] (fin2) at (3.3,-2.12) {};
	\node[dspnodefull] (finnode) at (6,-2) {};
	\node[box, minimum size=1cm,minimum width=2cm] (mru) at (6,-4) {\begin{tabular}{c}Model \&\\residual\\update (\ref{eq:modelUpdate})\end{tabular}};
	\node[box, minimum size=1cm,minimum width=2cm] (ce) at (6,-6) {\begin{tabular}{c}\!\!\!Initial coefficient\!\!\!\\estimation (\ref{eq:coefficient-estimation})\end{tabular}};
	\node[box, minimum size=1cm,minimum width=2cm] (cs) at (3.3,-6) {\begin{tabular}{c}Coefficient\\selection (\ref{eq:coeff-sel})\end{tabular}};
	\node[box, minimum size=1cm,minimum width=2cm] (ocu) at (3.3,-4) {\begin{tabular}{c}Optimal\\coefficient\\update (\ref{eq:coeffUpdate})\end{tabular}};s
	\node[] (out) at (2,-8) {\begin{tabular}{c}Optimized transformed signal $\boldsymbol{\tilde{S}}$\end{tabular}};
	
	\draw[->] (in1) -- (cca) node {};
	\draw[->] (in2) -- (bfd) node {};
	\draw[->] (bfd) -- (mru) node [pos=0.2,right] {$\mathcal{D}_t$};
	\draw[->] (cca) -- (fin) node[pos=0.5,right] {$N$};
	\draw[->] (fin) -| (mru) node [pos=0.25,above] {No};
	\draw[->] (mru) -- (ce) node {};
	\draw[->] (ce) -- (cs) node {};
	\draw[->] (cs) -- (ocu) node {};
	\draw[->] (ocu) -- (fin2) node {};
	\draw[->] (fin) -- (out) node [pos=0.5,left] {Yes};
	\end{tikzpicture}
	\vspace{-.5cm}
	\caption{Overview of the proposed ROSE method. The signal $\boldsymbol{s}$ is processed together with mask $\boldsymbol{m}$ denoting the occupied pixels and the applied transform to obtain a model, which is optimized only on the occupied samples.}
	\label{img:proposed-method}
\end{figure}
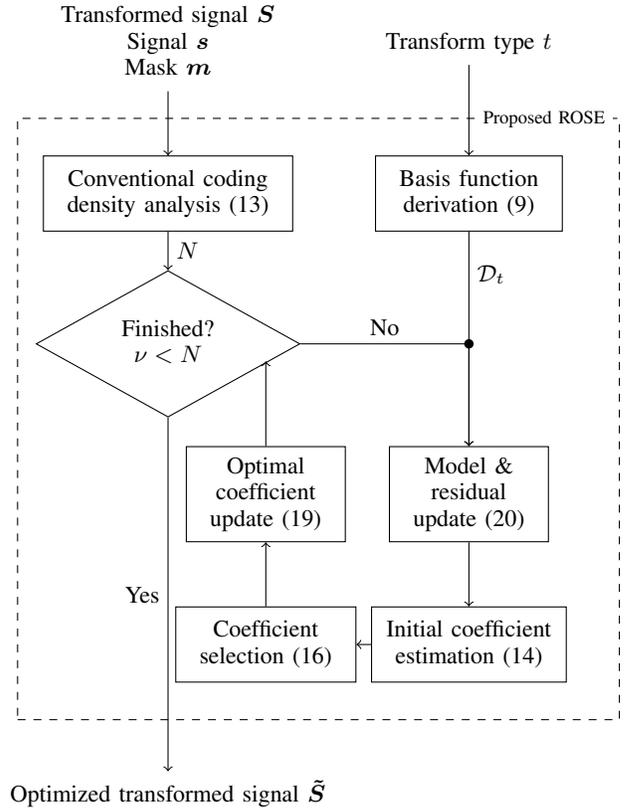
\revision{An overview on this method is shown in Fig.~\ref{img:proposed-method}. First, the share of nonzero coefficients is determined (coding density analysis) and basis functions are derived. Afterwards, a while-loop is launched (bottom right) that iteratively determines the optimal residual coefficients. }

\subsection{Rate-Distortion Control}
It is well known that in video compression, decisions should be based on rate-distortion costs in order to obtain optimal compression results \cite{Sullivan98}. In rate-distortion optimization (RDO), the coding costs $J$ are minimized by 
\begin{equation}
\min J = D + \lambda\cdot R, 
\end{equation}
where $D$ is the distortion, $R$ is the rate, and $\lambda$ is a Lagrange multiplier derived from the pre-defined QP. Accordingly, we interpret the model error  $\boldsymbol{e}_{\nu,k}$ from eq. \eqref{eq:modelError} as the distortion $D$ and select the optimal coefficient according to eq. \eqref{eq:coeff-sel}, 
where $\lambda$ is the same Lagrange multiplier as used in standard RDO and $b$ is the number of bits needed to code the current selection of residual coefficients including coefficient $k_\nu$. 

In order to determine the accurate number of bits $b$ for the current selection of residual coefficients, the coefficients must be encoded using the context-adaptive binary arithmetic coding (CABAC) engine \cite{CABAC}. Due to the large number of iterations and the resulting high complexity, we propose to simplify the determination of the number of bits by estimating $b$ using a rate model. For this, we consider two models from the literature. The first model is designed to estimate the cross-entropy loss \cite{Helle17} and is abbreviated by `{\small log}' in the following. We adapt it for the application of rate estimation as
\begin{equation}
\hat b_\mathrm{log}(\boldsymbol{\check c}) = \sum_{k\in \mathcal{\hat K}} \alpha\cdot |\boldsymbol{\check c}_k| + \beta g\left(\gamma |\boldsymbol{\check c}_k| - \delta\right), \label{eq:rateModel}
\end{equation}
where $g()$ is the logistic function 
\begin{equation}
g(x) = \frac{1}{1+\mathrm{e}^{-x}} 
\end{equation}
and the parameters $\alpha$, $\beta$, $\gamma$, and $\delta$ describe the relation between the coefficients' values and the true rate. 

Due to the definition of the function, we can see that in this model, only nonzero coefficients are considered in rate estimation. As such, this approach favors a low number of coefficients, which is in line with the goal of ROSE. Second, we can see that the value of the coefficient is considered with a linear term and with a logistic function. With this combination, the rate increment becomes smaller when the coefficient value increases. This behavior favors a choice of few coefficients with potentially highly variable values. 

The second model (`{\small stat}') was designed for VVC intra coding and is based on the coefficients' statistics \cite{Brand22}. It is defined as
\begin{equation}
\hat b_\mathrm{stat}(\boldsymbol{\check c}) = \alpha \cdot \left| \mathcal{\hat K}\right| + \beta \cdot L(\boldsymbol{\check c}) + \gamma\cdot Z + \delta\cdot E , 
\label{eq:modelFabi}
\end{equation}
where $L$ is the sum of the logarithms of the coefficient values
\begin{equation}
L(\boldsymbol{\check c})=\sum_{k\in \mathcal{\hat K}} \log_2(\left|\boldsymbol{\check c}_k\right|). 
\end{equation}
$Z$ is calculated by the sum of the last zig-zag-scan-coefficient indices of each $4\times 4$ subblock and $E$ is the sum of the subblocks' entropy calculated by 
\begin{equation}
H \left( \frac{N_1}{16} \right), 
\end{equation}
where the entropy is obtained using $N_1$, which is the number of coefficients strictly greater one. $H(p) = -p\log_2(p)-(1-p)\log_2(1-p)$ is the entropy function. 
The parameters $\alpha, \beta, \gamma$, and $\delta$ describe the relation between the block statistics and the rate. 

For training the parameters $\alpha$, $\beta$, $\gamma$, and $\delta$ of both models, \revision{we analyze the statistics during HM encoding of a single, random intra-coded frame (BasketballDrill from the JVET common test conditions \cite{JVET_N1010}) at a medium QP of 32} and extract the corresponding statistical information on the coefficients and the bits written by the CABAC encoder. The parameters are then trained by least-squares curve fitting. The resulting values, which are then used for our ROSE implementation, are listed in Table~\ref{tab:helleParams}. \revision{We report the mean relative estimation error of the two models given by 
\begin{equation}
\varepsilon = \frac{1}{\left|\mathcal{B}\right|}\sum_{B=1}^{\left|\mathcal{B}\right|} \frac{\left|\hat b_B - b_B\right|}{b_B},
\end{equation}
where $B$ is the block index of all blocks $\mathcal{B}$ used for training, $b_B$ is the true number of bits for block $B$, and $\hat b_B$ is the estimated number of bits for block $B$. We compare the estimation accuracy in three different cases: The training estimation error on the original data $\varepsilon_\mathrm{train}$, the validation error on geometry data $\varepsilon_\mathrm{geom}$, and the validation error on texture data $\varepsilon_\mathrm{text}$. The corresponding values shown in Table~\ref{tab:helleParams} show that the `{\small stat}' model returns a much lower estimation error than the `{\small log}' model in general, because the `{\small stat}' errors are always less than half of the `{\small log}' errors. For the `{\small stat}' model, mean errors of $26\%$ and $27\%$ can be observed for geometry and texture data, respectively.  }   In the following, the ROSE variant with the {\small `log'}-rate model is denoted ROSEL, the variant with the {\small `stat'}-rate model is denoted ROSES.

\begin{table}[t]
\renewcommand{\arraystretch}{1.3}
\caption{\revision{Trained values for the two bitrate models from eqs. \eqref{eq:rateModel} and \eqref{eq:modelFabi} (columns $\alpha$ to $\delta$). In the remaining columns on the right, the mean relative estimation errors of the bitrate models are reported for the training data (column $\varepsilon_\mathrm{train}$), the validation for geometry data (column $\varepsilon_\mathrm{geom}$), and the validation for texture data (column $\varepsilon_\mathrm{text}$). } }
\vspace{-.5cm}
\label{tab:helleParams}
\begin{center}
\resizebox{.48\textwidth}{!}{
\begin{tabular}{l||r|r|r|r||r|r|r}
\hline
Model & $\alpha$ & $\beta$ & $\gamma$ & $\delta$  & $\varepsilon_\mathrm{train}$ & $\varepsilon_\mathrm{geom}$ & $\varepsilon_\mathrm{text}$\\
\hline
log \cite{Helle17} & $2.410$  & $4.425$ & $0.036$ & $9.427$  & $28\%$& $54\%$& $55\%$ \\
stat \cite{Brand22} &  $1.096$ & $1.747$ & $6.275$ & $1.346$ & $12\%$& $26\%$& $27\%$\\
\hline
 \end{tabular}}
 \vspace{-0.5cm}
 \end{center}
\end{table}

\subsection{Fast ROSE}
The previously formulated ROSE algorithm offers the flexibility to use masks $\boldsymbol{m}$ with arbitrary weights $\boldsymbol{m}_i \in [0,1]$ so that more important pixels could be weighted stronger than other samples. However, for the investigated applications in this paper, all relevant pixels are assumed to be of equal importance and so, the mask is kept binary with $\boldsymbol{m}_i \in \{0,1\}$. In contrast to state-of-the-art signal extrapolation, this constraint offers a great potential for complexity optimization, which results in the fast ROSE. As the generation of basis functions must only be conducted once and can be stored in look-up tables, only the iterative algorithm is further analyzed in the following. As in coding of unoccupied regions, the goal is to estimate a model at positions where the mask is different from zero. Thus, the model estimation can be simplified to
\begin{align}
	\boldsymbol{g}^{\boldsymbol{m}} = \sum_{k \in \mathcal{\hat{K}}_t} \hat{c}_k \boldsymbol{\varphi}_{k}^{\boldsymbol{m}}
\end{align}
with superscript ${\boldsymbol{m}}$ denoting that the vector only contains the relevant entries with $\boldsymbol{m}_i = 1$. As a result, the number of pixels and the computational complexity can be significantly reduced especially for large unoccupied regions. The preliminary coefficient estimation from eq. (\ref{eq:coefficient-estimation}) is reformulated to
\begin{align}
	\boldsymbol{\check{c}}_{\nu,k} = \dfrac{\sum\limits_{i} \boldsymbol{r}^{\boldsymbol{m}}_{\nu,i} \boldsymbol{\varphi}^{\boldsymbol{m}}_{k,i}}{\sum\limits_{i} (\boldsymbol{\varphi}^{\boldsymbol{m}}_{k,i} )^2 }\text{ , } \forall k \text{ , }
	\label{eq:coeff-fose}
\end{align}
accordingly. Apparently, the term $\sum\limits_{i} (\boldsymbol{\varphi}^{\boldsymbol{m}}_{k,i} )^2$
is independent of the signal itself and can be precalculated in a look-up table on block level, so that only the simplified numerator has to be evaluated during the iterative procedure. Next, the basis function selection is reformulated to
\begin{align}
	\hat{k}_{\nu} = \argmin\limits_{k} \left\{ | \boldsymbol{r}^{\boldsymbol{m}}_{\nu-1,i} - 
	\boldsymbol{\check{c}}_{\nu,k} 
	\boldsymbol{\varphi}^{\boldsymbol{m}}_{k,i} |^2 + \lambda \cdot b(\boldsymbol{\check c}) \right\} \mathrm{\ .}
	\label{eq:coeff-sel-fose}
\end{align}
Furthermore, the coefficient estimation simplifies to
\begin{align}
	\begin{pmatrix}
		\boldsymbol{\hat{c}}^{\boldsymbol{m}}_{0}\\
		\vdots\\
		\boldsymbol{\hat{c}}^{\boldsymbol{m}}_{\nu}
	\end{pmatrix}
	=
	\left(
	\begin{bmatrix}
	\boldsymbol{\varphi}^{\boldsymbol{m}}_{\hat{k}_{0}}\\ \vdots\\ \boldsymbol{\varphi}^{\boldsymbol{m}}_{\hat{k}_{\nu}}
	\end{bmatrix}
	\begin{bmatrix}
	\boldsymbol{\varphi}^{\boldsymbol{m}}_{\hat{k}_{0}}\\ \vdots\\ \boldsymbol{\varphi}^{\boldsymbol{m}}_{\hat{k}_{\nu}}
	\end{bmatrix}^\intercal
	\right)^{-1}
	\begin{bmatrix}
	\boldsymbol{\varphi}^{\boldsymbol{m}}_{\hat{k}_{0}}\\ \vdots\\ \boldsymbol{\varphi}^{\boldsymbol{m}}_{\hat{k}_{\nu}}
	\end{bmatrix}
	\boldsymbol{s}^{\boldsymbol{m}}
	\text{ .}
	\label{eq:opt-coeff-fose}
\end{align}
Besides smaller vector lengths, also the number of matrix multiplications is considerably reduced. Then, the model is obtained as
\begin{align}
	\boldsymbol{g}^{\boldsymbol{m}}_{\nu} =
	\begin{bmatrix}
		\boldsymbol{\varphi}^{\boldsymbol{m}}_{\hat{k}_{0}}\\
		\vdots\\
		\boldsymbol{\varphi}^{\boldsymbol{m}}_{\hat{k}_{\nu}}
	\end{bmatrix}
	\begin{pmatrix}
		\boldsymbol{\hat{c}}_{0}\\
		\vdots\\
		\boldsymbol{\hat{c}}_{\nu}
	\end{pmatrix} \text{ ,}
	\label{eq:model-update}
\end{align}
and the residual update results in
\begin{align}
	\boldsymbol{r}_{\nu+1}^{\boldsymbol{m}} = \boldsymbol{s}^{\boldsymbol{m}} - \boldsymbol{g}^{\boldsymbol{m}}_{\nu} \text{ .}
\end{align}
Depending on the actual application and the number of unoccupied pixels, the speed-up can range to several orders of magnitude. Nevertheless, also for only few unoccupied samples, the fast ROSE approach gives a faster computation than general ROSE without any loss of coding performance due to fewer matrix multiplications and the large number of precomputed calculations.

\section{Evaluation}
\label{sec:eval}

In this chapter, the compression performance of the proposed fast ROSE algorithm is evaluated in detail. To this end, Subsection~\ref{sec:evalSetup} presents the input data set and gives details on the evaluation setup. Afterwards, the compression results are discussed. 

\begin{figure*}[t]
\centering
\includegraphics[width=.99\textwidth]{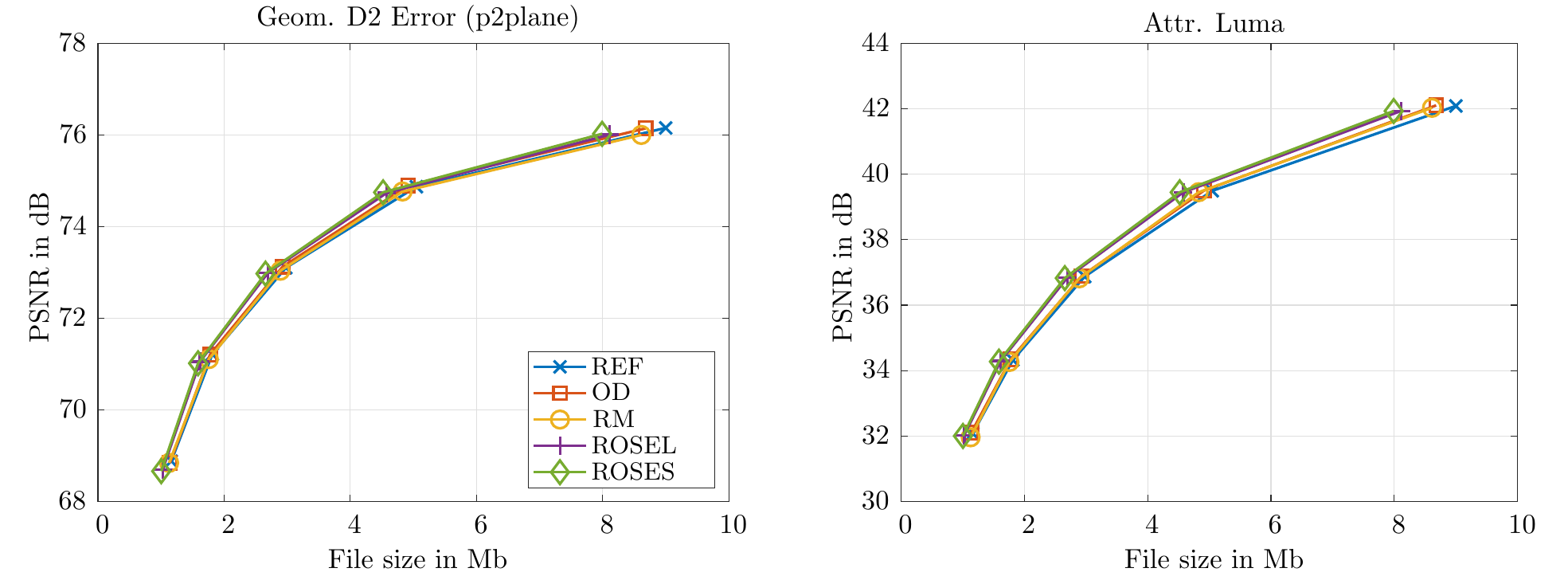}
\vspace{-.3cm} 
\caption{RD-performance of different encoding algorithms exploiting unoccupied regions for the Loot point cloud. The encoder configuration is all-intra. The vertical axes show the error metrics D2 error (left) and the Y-PSNR (right). The colors represent the reference encoder in blue (REF), the occupied distortion calculation in red (OD), the algorithm with residual masking (RM) in yellow, ROSE with rate estimation based on the logistic function (ROSEL) in purple, and ROSE with rate estimation based on coefficient statistics (ROSES) in green.  } 
\label{fig:RD-PCC}
\end{figure*}

\subsection{Evaluation Setup}
\label{sec:evalSetup}
For the evaluation of the proposed ROSE algorithm, we use point clouds from the MPEG common test conditions \cite{Schwarz18}. The point clouds are listed in Table~\ref{tab:pccSet}. In order to reduce the simulation time, we encode 16 frames from each point cloud in all-intra and randomaccess configuration. For the occupancy map, the standard lossless encoder is used, for the geometry map and the texture map, the proposed encoder is used. As an encoding framework, we use TMC2-8.0 with the HM encoder version 16.20{\_}SCM8.8, which includes the standard-conform extensions proposed in \cite{Li20}. 

\revision{
 Note that in this work, the proposed ROSE is restricted to intra-coded blocks because of the following reasons. First, the overall number of nonzero coefficients is significantly smaller in inter-predicted blocks, such that also potential bitrate savings are smaller. Second, the used rate models are designed for intra-predicted blocks \cite{Brand22} and in our tests, we found that the estimation accuracy decreases drastically when using inter-predicted blocks. As a consequence, ROSE often leads to suboptimal coefficient selections such that rate savings are lost, as indicated by our experiments. }

For encoding, we use the five RD-points proposed in \cite{Schwarz18}. The resulting bitrates and distortions are reported for the original encoding procedure using pixel padding (REF), the approach using distortion calculation for occupied regions proposed in \cite{Li20} (OD), the masking of residual signals proposed in \cite{Herglotz19c} (RM), 
and the fast ROSE approach with coefficient selection based on the estimated RD costs for both the `log'-rate model (ROSEL) and the `stat'-rate model (ROSES) from eqs. \eqref{eq:rateModel} and \eqref{eq:modelFabi}, respectively. 

We evaluate our results using the quality metrics proposed in \cite{Schwarz18,Tian16}, which are the point-to-point error (D1) and the point-to-plane error (D2) for the geometry map. For the texture map, we use the Y-PSNR, the U-PSNR, and the V-PSNR. \revision{To assess the rate, we perform comparisons with the overall bitrate and the separate geometry as well as texture bitrates.} Furthermore, we report average rate savings using the well-known Bj{\o}ntegaard-Delta rate (BD-rate) from \cite{Bjonte01} using piecewise cubic interpolation \cite{Strom21}. 

\begin{table}[t]
\renewcommand{\arraystretch}{1.3}
\caption{Point clouds for the evaluation of the proposed algorithm taken from \cite{Schwarz18}.}
\vspace{-.5cm}
\label{tab:pccSet}
\begin{center}
\begin{tabular}{l|l|l}
\hline
Name & fps & Number of Points per frame\\
\hline
Queen & 50 & ${\sim} 1{,}000{,}000$\\
Loot & 30 & ${\sim} 780{,}000$\\
Longdress & 30 & ${\sim} 800{,}000$\\
Red and Black & 30 &$ {\sim} 700{,}000$\\
Soldier & 30 & ${\sim} 1{,}000{,}000$\\
\hline
 \end{tabular}
 \end{center}
\end{table}

\revision{Note that the proposed ROSE algorithm only processes blocks that include both occupied and unoccupied pixels. As the block partitioning and hence the block sizes depend on RDO decisions, the overall share of pixels that is affected by ROSE depends on encoder decisions. Therefore, we calculated the potential share of blocks affected by ROSE depending on fixed block sizes. For the first frame of the Loot point cloud, assuming that encoding is performed with a fixed block size, the corresponding shares are 
\begin{itemize}
\item $8\times 8$-blocks: $5.7\%$,
\item $16\times 16$-blocks: $16.5\%$, 
\item $32\times 32$-blocks: $36\%$. 
\end{itemize} 
The share for $4\times 4$-blocks is zero because of the occupancy map's definition based on $4\times 4$ blocks. We can see that a maximum of $36\%$ of the frame's pixels is affected by ROSE. }

\subsection{Rate-Distortion Performance}

Fig.~\ref{fig:RD-PCC} shows exemplary RD-curves for the Loot point cloud encoded in the all-intra configuration, where we show  the geometry D2 error (left) and the Y-PSNR of the texture (right). For the representation of the bitrate, we use the overall file size after compression.

We can see that the proposed ROSES (green lines) shows highest compression ratios because for all tested bitrates, the curve is located on the left of the other curves. This means that at the same visual quality, a lower bitrate is needed. Similar curves can be observed for the other error metrics (D1 error, U-PSNR, and V-PSNR). 

In order to obtain average values, we report the compression performance in terms of the BD-rate \cite{Schwarz18} and show results for geometry metrics in Table~\ref{tab:BDR_pcc_geom} and results for texture metrics in Table~\ref{tab:BDR_pcc_texture}. The results are given for OD from \cite{Li20}, RM from \cite{Herglotz19c}, ROSEL, 
and ROSES. 
All values are calculated with respect to reference encoding (REF). 
\begin{table*}[ht]
\renewcommand{\arraystretch}{1.3}
\caption{BD-rate values for geometry error metrics and all tested sequences in the all-intra and in the randomaccess configuration.  }
\label{tab:BDR_pcc_geom}
\vspace{-.3cm}
\begin{center}
\resizebox{.7\textwidth}{!}{ 
\begin{tabular}{l||r|r|r|r||r|r|r|r}
\hline

\multicolumn{9}{c}{\textbf{all-intra}}\\ \hline
& \multicolumn{4}{c||}{Geometry BD-rate (D1)} & \multicolumn{4}{c}{Geometry BD-rate (D2)}\\
& OD & RM & ROSEL & ROSES & OD & RM & ROSEL & ROSES\\ \hline
Loot
 & $-2.36\%$
 & $-0.53\%$
 & $-6.82\%$
 & $-8.1\%$
 & $-2.34\%$
 & $-0.73\%$
 & $-6.58\%$
 & $-7.8\%$
\\
Redandblack
 & $-1.67\%$
 & $0.08\%$
 & $-3.23\%$
 & $-3.9\%$
 & $-1.85\%$
 & $0.13\%$
 & $-3.23\%$
 & $-3.8\%$
\\
Soldier
 & $-1.89\%$
 & $3.34\%$
 & $-2.7\%$
 & $-3.6\%$
 & $-1.98\%$
 & $2.98\%$
 & $-2.53\%$
 & $-3.2\%$
\\
Queen
 & $-1.7\%$
 & $2.32\%$
 & $-6.56\%$
 & $-7.8\%$
 & $-1.82\%$
 & $2.12\%$
 & $-5.79\%$
 & $-6.8\%$
\\
Longdress
 & $-1.44\%$
 & $0.61\%$
 & $-0.97\%$
 & $-0.8\%$
 & $-1.36\%$
 & $0.34\%$
 & $-0.72\%$
 & $-0.5\%$
\\
\hline
Average
 & $-1.81\%$
 & $1.16\%$
 & $-4.06\%$
 & $\mathbf{-4.84\%}$
 & $-1.87\%$
 & $0.97\%$
 & $-3.77\%$
 & $\mathbf{-4.42\%}$
\\
\hline\hline
\multicolumn{9}{c}{\textbf{randomaccess}}\\ \hline
& \multicolumn{4}{c||}{Geometry BD-rate (D1)} & \multicolumn{4}{c}{Geometry BD-rate (D2)}\\
& OD & RM & ROSEL & ROSES & OD & RM & ROSEL & ROSES\\ \hline
Loot
 & $-21.69\%$
 & $-22.68\%$
 & $-23.57\%$
 & $-24.55\%$
 & $-21.75\%$
 & $-22.34\%$
 & $-23.28\%$
 & $-24.24\%$
\\
Redandblack
 & $-11.11\%$
 & $-11.01\%$
 & $-11.72\%$
 & $-12.69\%$
 & $-11.29\%$
 & $-11.12\%$
 & $-11.71\%$
 & $-12.54\%$
\\
Soldier
 & $-20.18\%$
 & $-17.11\%$
 & $-19.63\%$
 & $-20.25\%$
 & $-19.97\%$
 & $-17.13\%$
 & $-19.3\%$
 & $-19.86\%$
\\
Queen
 & $-17.99\%$
 & $-15.63\%$
 & $-19.32\%$
 & $-19.23\%$
 & $-18.22\%$
 & $-15.59\%$
 & $-18.8\%$
 & $-18.46\%$
\\
Longdress
 & $-8.62\%$
 & $-7.04\%$
 & $-8.17\%$
 & $-8.25\%$
 & $-8.93\%$
 & $-7.63\%$
 & $-7.98\%$
 & $-8.35\%$
\\
\hline
Average
 & $-15.92\%$
 & $-14.69\%$
 & $-16.48\%$
 & $\mathbf{-16.99\%}$
 & $-16.03\%$
 & $-14.76\%$
 & $-16.21\%$
 & $\mathbf{-16.69\%}$
\\
\hline
 \end{tabular}}
 \end{center}
 \vspace{-.5cm}
\end{table*}

\begin{table*}[t]
\renewcommand{\arraystretch}{1.3}
\caption{BD-rate values for texture error metrics and all tested sequences in the all-intra and in the randomaccess configuration.  }
\label{tab:BDR_pcc_texture}
\vspace{-.7cm}
\begin{center}
\resizebox{\textwidth}{!}{ 
\begin{tabular}{l||r|r|r|r||r|r|r|r||r|r|r|r}
\hline

\multicolumn{13}{c}{\textbf{all-intra}}\\ \hline
& \multicolumn{4}{c||}{Texture BD-rate (Y)} & \multicolumn{4}{c||}{Texture BD-rate (U)} & \multicolumn{4}{c}{Texture BD-rate (V)}\\
& OD & RM & ROSEL & ROSES & OD & RM & ROSEL & ROSES& OD & RM & ROSEL & ROSES \\ \hline
Loot
 & $-2.58\%$
 & $-2.4\%$
 & $-9.12\%$
 & $-10.3\%$
 & $-2.01\%$
 & $0.15\%$
 & $-8.32\%$
 & $-10.7\%$
 & $-1.65\%$
 & $0.58\%$
 & $-8.65\%$
 & $-9.9\%$
\\
Redandblack
 & $-2.24\%$
 & $-0.68\%$
 & $-5.89\%$
 & $-6.9\%$
 & $-1.85\%$
 & $-1.34\%$
 & $-7.35\%$
 & $-8.8\%$
 & $-2.13\%$
 & $-1.86\%$
 & $-7.52\%$
 & $-8.9\%$
\\
Soldier
 & $-2.2\%$
 & $0.96\%$
 & $-5.73\%$
 & $-6.8\%$
 & $-0.34\%$
 & $5.96\%$
 & $-1.44\%$
 & $-3.4\%$
 & $-1.76\%$
 & $5.9\%$
 & $-1.84\%$
 & $-3.9\%$
\\
Queen
 & $-1.62\%$
 & $-0.79\%$
 & $-5.14\%$
 & $-6.6\%$
 & $-2.79\%$
 & $-1.8\%$
 & $-9.49\%$
 & $-11.2\%$
 & $-2.14\%$
 & $-1\%$
 & $-8.18\%$
 & $-10.1\%$
\\
Longdress
 & $-1.15\%$
 & $-0.3\%$
 & $-2.92\%$
 & $-3.3\%$
 & $-1.3\%$
 & $-1.09\%$
 & $-4.39\%$
 & $-5\%$
 & $-1.37\%$
 & $-0.99\%$
 & $-4.07\%$
 & $-4.7\%$
\\
\hline
Average
 & $-1.96\%$
 & $-0.64\%$
 & $-5.76\%$
 & $\mathbf{-6.78\%}$
 & $-1.66\%$
 & $0.38\%$
 & $-6.20\%$
 & $\mathbf{-7.82\%}$
 & $-1.81\%$
 & $0.53\%$
 & $-6.05\%$
 & $\mathbf{-7.50\%}$
\\
\hline\hline
\multicolumn{13}{c}{\textbf{randomaccess}}\\ \hline
& \multicolumn{4}{c||}{Texture BD-rate (Y)} & \multicolumn{4}{c||}{Texture BD-rate (U)} & \multicolumn{4}{c}{Texture BD-rate (V)}\\
& OD & RM & ROSEL & ROSES & OD & RM & ROSEL & ROSES& OD & RM & ROSEL & ROSES \\ \hline
Loot
 & $-21.41\%$
 & $-20.17\%$
 & $-21.84\%$
 & $-23.15\%$
 & $-20.97\%$
 & $-18.65\%$
 & $-20.56\%$
 & $-22.84\%$
 & $-22.4\%$
 & $-20\%$
 & $-21.61\%$
 & $-24.19\%$
\\
Redandblack
 & $-12.16\%$
 & $-10.95\%$
 & $-11.9\%$
 & $-13.25\%$
 & $-11.21\%$
 & $-10.94\%$
 & $-11.71\%$
 & $-13.97\%$
 & $-12.86\%$
 & $-12.24\%$
 & $-13.73\%$
 & $-15.28\%$
\\
Soldier
 & $-23.35\%$
 & $-19.21\%$
 & $-21.45\%$
 & $-22.58\%$
 & $-21.07\%$
 & $-15.49\%$
 & $-19.97\%$
 & $-19.73\%$
 & $-22.4\%$
 & $-15.04\%$
 & $-18.36\%$
 & $-20.28\%$
\\
Queen
 & $-19.5\%$
 & $-16.54\%$
 & $-19.41\%$
 & $-19.92\%$
 & $-19.46\%$
 & $-17.08\%$
 & $-19.76\%$
 & $-21.07\%$
 & $-19.06\%$
 & $-16.8\%$
 & $-20.19\%$
 & $-20.12\%$
\\
Longdress
 & $-10.04\%$
 & $-8.71\%$
 & $-9.56\%$
 & $-10.05\%$
 & $-9.9\%$
 & $-8.87\%$
 & $-9.84\%$
 & $-10.88\%$
 & $-9.87\%$
 & $-8.95\%$
 & $-9.87\%$
 & $-10.77\%$
\\
\hline
Average
 & $-17.29\%$
 & $-15.12\%$
 & $-16.83\%$
 & $\mathbf{-17.79\%}$
 & $-16.52\%$
 & $-14.21\%$
 & $-16.37\%$
 & $\mathbf{-17.70\%}$
 & $-17.32\%$
 & $-14.61\%$
 & $-16.75\%$
 & $\mathbf{-18.13\%}$
\\
\hline

\hline

 \end{tabular}}
 \end{center}
\end{table*}

Considering the all-intra configuration, we find that ROSES outperforms all the other configurations. In particular, rate savings for all error metrics yield more than $4\%$ \revision{for geometry and $6\%$ for texture metrics, respectively}, where state-of-the-art algorithms (OD and RM) achieve less than $2\%$ of savings \revision{(see Tables~\ref{tab:BDR_pcc_geom} and \ref{tab:BDR_pcc_texture})}. 

Comparing the rate savings for different bitrate estimators (ROSEL and ROSES), we find that the statistic based estimator leads to higher bitrate savings in general (average savings are always more than $0.5\%$ higher). This \revision{confirms} that the more sophisticated rate model returns more reliable rate estimates than the other rate model. Furthermore, we can see that the proposed ROSE performs solidly because an enhanced rate model directly leads to additional bitrate savings. 
\par
Considering the randomaccess case (\revision{bottom of Tables~\ref{tab:BDR_pcc_geom} and \ref{tab:BDR_pcc_texture})}, rate-savings are also significant. For all error metrics, ROSES outperforms the second best method from the literature (OD) by more than $0.5\%$. However, these additional savings are lower than in the all-intra case. The reason is that in the proposed algorithm, ROSE is only applied to intra-predicted blocks, which occur less frequently in the randomaccess case, such that potential bitrate savings are lower. \revision{Still, it is worth mentioning that the overall bitrate savings with respect to all-intra coding are much higher (more than $15\%$). These large savings are mainly caused by the OD algorithm proposed in \cite{Li20}, which uses a dedicated motion estimation technique in encoding. }

\revision{For a more in-depth analysis, we also report average BD-rate savings when considering the separate geometry and texture bitrates. Note that the occupancy-map bitrates did not change for all tested algorithms. 	The results are listed in Table~\ref{tab:BDR_pcc_partRates}. }
\begin{table}[ht]
\renewcommand{\arraystretch}{1.3}
\caption{\revision{BD-rate values considering only the geometry rate (D1, D2) and the attribute rate (Y, U, V). The values are averaged over all sequences. } }
\label{tab:BDR_pcc_partRates}
\vspace{-.5cm}
\begin{center}
\begin{tabular}{l||r|r|r|r}
\hline

\multicolumn{5}{c}{\textbf{all-intra}}\\ \hline
& OD & RM & ROSEL & ROSES \\ \hline
D1 (geom. rate)
 & $-3.8\%$
 & $4.1\%$
 & $-8.3\%$
 & $\mathbf{-10.5\%}$
\\
D2 (geom. rate)
 & $-3.9\%$
 & $4\%$
 & $-8.2\%$
 & $\mathbf{-10.3\%}$
\\
Y (attr. rate)
 & $-1\%$
 & $-2.6\%$
 & $-4.2\%$
 & $\mathbf{-4.7\%}$
\\
U (attr. rate)
 & $-0.6\%$
 & $-0.8\%$
 & $-4.5\%$
 & $\mathbf{-5.7\%}$
\\
V (attr. rate)
 & $-0.8\%$
 & $-0.7\%$
 & $-4.3\%$
 & $\mathbf{-5.3\%}$
\\
\hline\hline
\multicolumn{5}{c}{\textbf{randomaccess}}\\ \hline
& OD & RM & ROSEL & ROSES \\ \hline
D1 (geom. rate)
 & $-17.5\%$
 & $-16\%$
 & $-18.6\%$
 & $\mathbf{-19.7\%}$
\\
D2 (geom. rate)
 & $-17.6\%$
 & $-16.1\%$
 & $-18.6\%$
 & $\mathbf{-19.6\%}$
\\
Y (attr. rate)
 & $\mathbf{-15.6\%}$
 & $-13.2\%$
 & $-13.8\%$
 & $-14.3\%$
\\
U (attr. rate)
 & $\mathbf{-14.6\%}$
 & $-11.8\%$
 & $-13.2\%$
 & $-14.2\%$
\\
V (attr. rate)
 & $\mathbf{-15.8\%}$
 & $-12.4\%$
 & $-13.8\%$
 & $-14.9\%$
\\
\hline\hline
\end{tabular}
\end{center}
\end{table}
\revision{Considering the all-intra case, we can see that the proposed ROSE outperforms all state-of-the-art algorithms for all error metrics. In particular, BD-rate savings for the geometry map are highest with BD-rate values greater than $10\%$. Also concerning the randomaccess case, highest rate savings close to $20\%$ can be observed for the geometry errors. In contrast, texture rate savings reported for OD in the randomaccess case are slightly below savings reported for ROSES, which shows that when allowing inter coding, the accuracy of the rate model is insufficient. Still, as was shown in Table~\ref{tab:BDR_pcc_texture}, rate savings from geometry coding outweigh savings from texture coding because the overall bitrate could be reduced. }

\begin{figure}[t]
\centering
\includegraphics[width=.49\textwidth]{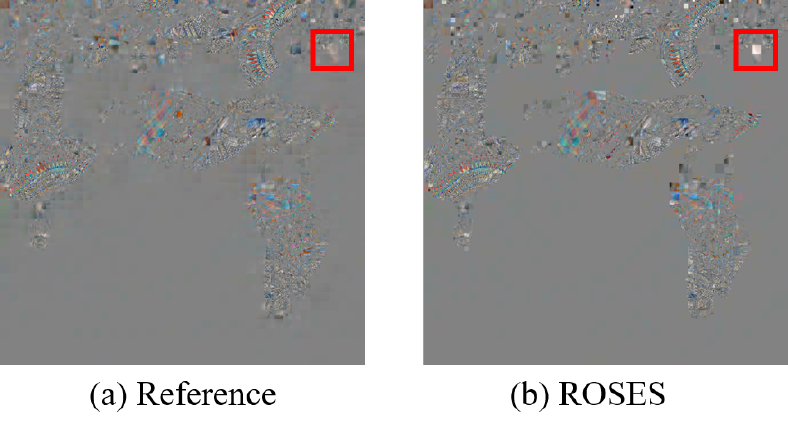} 
\vspace{-1cm}
\caption{\revision{Visualization of the coded residuals for the first frame of the Longdress point cloud, texture map, rate point r3. Grey corresponds to zero, the mapping is logarithmic to enhance visibility.  }}
\label{fig:visExampleLongdressResi}
\end{figure}

\revision{To illustrate the cause for the observed rate savings in intra coding, we compare the residual signal of an examplary intra-coded frame. Therefore, we plot the luma residual of the Longdress point cloud encoded at rate point r3 for the reference algorithm and ROSES in Fig.~\ref{fig:visExampleLongdressResi}. A residual of zero is represented by a grey color, bright colors represent a positive residual, and dark colors a negative residual. For better visibility, the brightness is mapped logarithmically. }

\revision{First of all, in both plots, one can easily locate the occupied regions because the residual is larger (high textural details), whereas the unoccupied regions are much smoother (especially bottom left). However, the boundaries between occupied and unoccupied regions are more pronounced in the plot representing ROSES (Fig.~\ref{fig:visExampleLongdressResi} (b)). Considering the unoccupied regions close to the boundaries, one can see that the amount of information, i.e. the variablility of the grey color, is much larger in the reference plot than in the ROSES plot. This leads to parts of the reported bitrate savings. }

\revision{Second, we can see that several blocks with both occupied and unoccupied pixels show a much higher residual for ROSES (see, e.g., the indication by the red square). These blocks were encoded with a large DC-coefficient. As such, these blocks correspond to the effect of energy decompaction illustrated in Fig.~\ref{img:energy-decompaction}, which leads to bitrate savings. }

\revision{It is worth mentioning that in unoccupied blocks, RM shows the same behavior as ROSES, i.e. that the grey areas are smooth with only zero coefficients. However, as reported in Tables~\ref{tab:BDR_pcc_geom} and \ref{tab:BDR_pcc_texture}, this does not lead to overall bitrate savings (average bitrate increase is positive except for the texture (Y) with a BD-rate of $-0.65\%$). As a consequence, we conclude that RM only leads to savings in combination with ROSES performed on mixed blocks, because otherwise, rate increases due to energy-decompaction will outweigh the bitrate savings.} 

\subsection{Visual Comparisons}

\begin{figure*}[h!t]
\centering
\includegraphics[width=.99\textwidth]{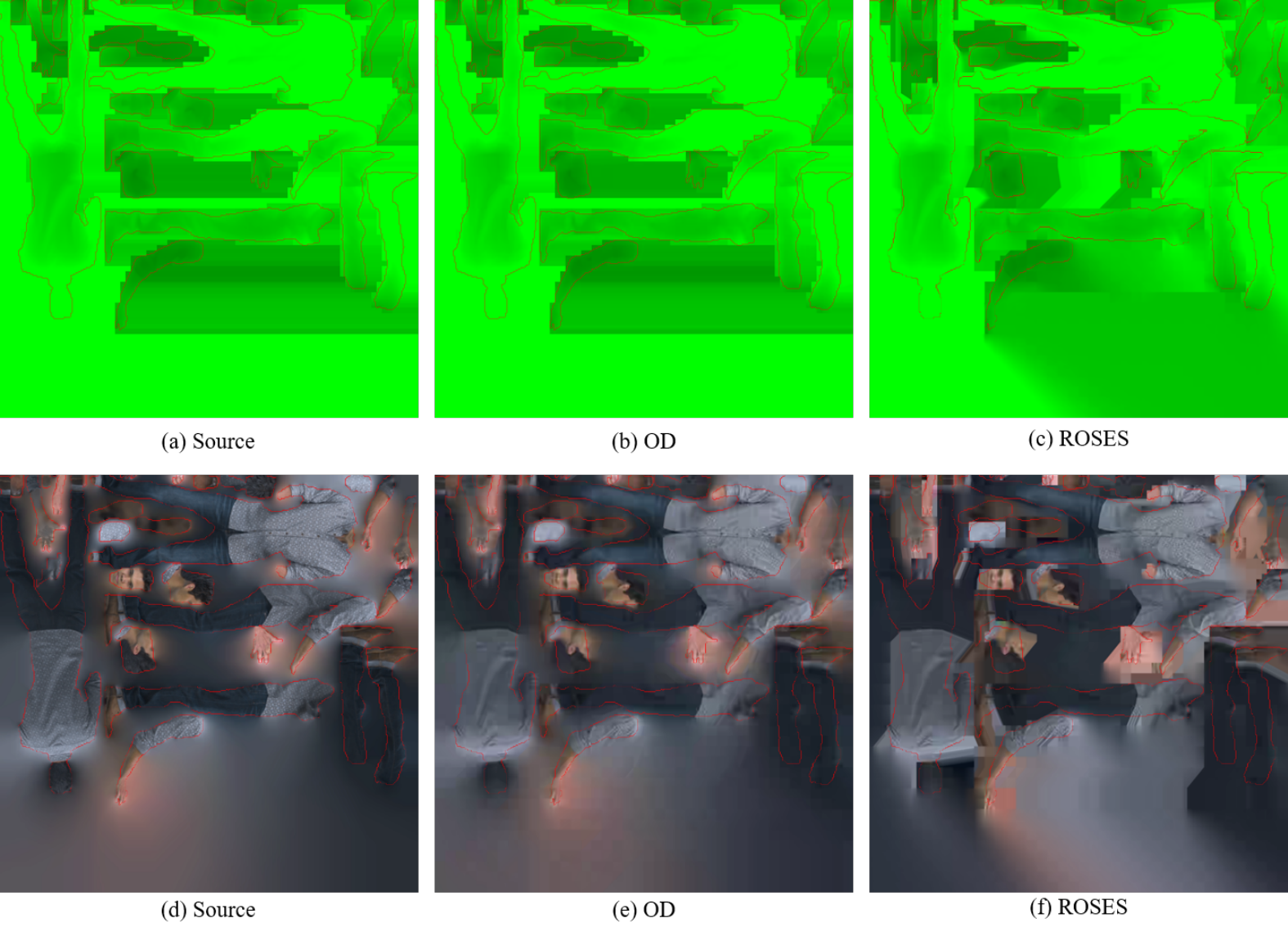} 
\vspace{-.5cm}
\caption{\revision{Geometry map (top) and texture map (bottom) for the first frame of the Loot point cloud. The occupied regions are inside the red contours. (a) and (d) show the original source frame, (b) and (e) the reconstructed frame after OD encoding, and (c) and (f) the reconstructed frame after ROSES encoding at rate point 1. }}
\label{fig:visExampleLootComplete}
\end{figure*}

\revision{
To illustrate the visual effect for a subset of the tested algorithms, we compare selected reconstructed frames. Fig.~\ref{fig:visExampleLootComplete} shows the source frame (a),(d) and two reconstructed frames for OD (b),(e) and ROSES (c),(f) for the geometry map (top) and the texture map (bottom) with boundaries between occupied and unoccupied regions in red.  }

\revision{Concerning both the geometry and the texture maps, we can see that the visual appearances by OD (b),(e) and of the source (a),(d) are similar. In particular, distortions are only visible in the texture map (e) (e.g., there is some blocking in the unoccupied region at the bottom and the Loot's shirt is more blurry in OD). }

\revision{Regarding ROSES (c),(f), we can see that the occupied regions show similar artifacts (blurriness) as the corresponding regions in the OD image (b),(e). However, there are large distortions in the unoccupied regions in both the geometry and the texture map. These correspond to the homogeneous grey regions in the residual image discussed in Fig.~\ref{fig:visExampleLongdressResi}, which lead to significant bitrate savings. The observable gradients and textures in this regions are caused by the selected intra prediction modes. }

\begin{figure*}[h!t]
\centering
\includegraphics[width=.99\textwidth]{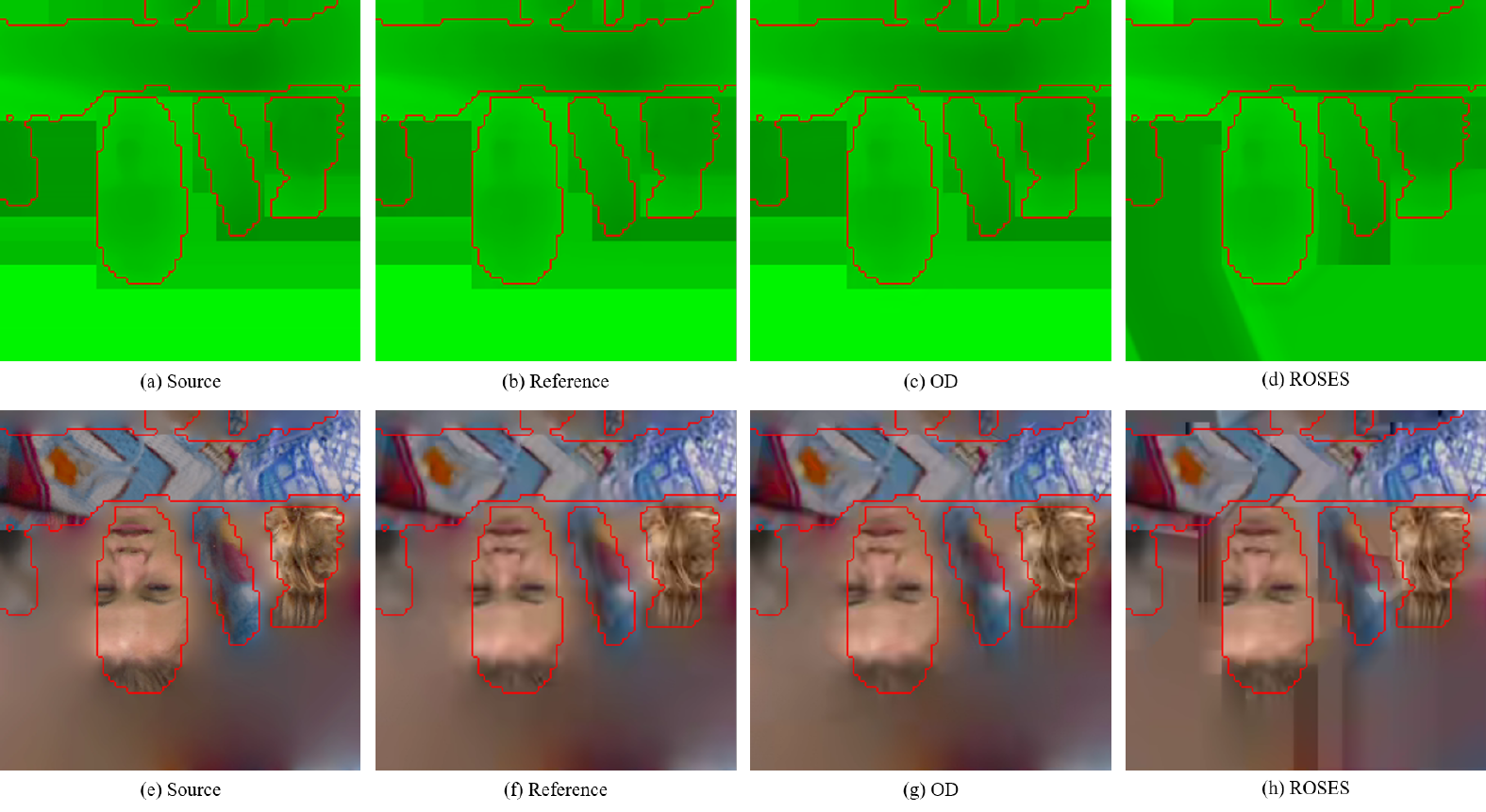} 
\vspace{-.5cm}
\caption{\revision{Close-ups of the geometry map (top) and the texture map (bottom) from the first frame of the Longdress point cloud (rate point 3). (a) and (e) show the input images, (b) and (f) the images after compression with the standard encoder, (c) and (g) the images with OD from \cite{Li20}, and (d) and (h) the images from the proposed ROSES encoding method. }  }
\label{fig:visExampleLongdressHead}
\end{figure*}

\revision{Finally, for a more detailed comparison, we show a zoom of a part of the reconstructed Longdress point cloud in Fig.~\ref{fig:visExampleLongdressHead}. The boundaries between occupied and unoccupied regions are indicated by red contours. }

\revision{Next to the large distortions in the unoccupied regions, the close-up allows discussions on distortions in occupied regions. One can see that distortions in occupied regions are similar for all algorithms and both geometry (b) to (d) as well as texture maps (f) to (h). In contrast, siginificant differences can be found at the borders of the occupied regions on the unoccupied side. For example, at the left of the woman's face, the extrapolation of the skin color  and the structure of the eye vanishes when using ROSES (h). Also, smooth color gradients between occupied regions are often replaced by homogeneous colors within coding blocks and strong edges at the coding block borders in unoccupied regions (e.g., right of the face). }

\subsection{\revision{Complexity}}
\revision{The complexity increases in terms of encoder processing time imposed by the tested algorithms, with ROSEL and ROSES based on the fast ROSE implementation, are reported in Table~\ref{tab:BDR_pcc_complexity}. Concerning all-intra coding, we can see that the complexity increases by a substantial factor of roughly $140$ on average. The reason is that the proposed algorithm performs an iterative searching procedure on every tested coding mode. However, one can see that the increase in complexity highly depends on the rate point and is lowest for the rate point targeting the lowest bitrate (r1). This can be explained by the restriction of ROSE, where the maximum number of iterations corresponds to the number of nonzero coefficients returned by regular transformation and quantization. As the overall number of nonzero coefficients increases with a lower QP, consequently also the number of iterations increases. }

\revision{Concerning randomacces, we can see that the increase in complexity is still significant (roughly $30$ on average), but much smaller than in the all-intra case. The reason is that due to early encoding decisions (e.g., early skip), the number of tested intra modes is much smaller.} 
\begin{table}[t]
\renewcommand{\arraystretch}{1.3}
\caption{\revision{Complexity comparison of all tested algorithms with standard encoding. The value represents the CPU time increase ($1$ means equal complexity). The CPU is an Intel(R) Core(TM) i7-8700 CPU @ 3.20GHz. The values are reported for different rate points and averaged over all tested sequences.   }}
\label{tab:BDR_pcc_complexity}
\vspace{-.5cm}
\begin{center}
\begin{tabular}{l||r|r|r|r}
\hline
\multicolumn{5}{c}{\textbf{all-intra}}\\ \hline
& OD & RM & ROSEL & ROSES \\ \hline
r1
 & $1.01$
 & $1.87$
 & $62.88$
 & $74.52$
\\
r2
 & $1.04$
 & $1.83$
 & $83.53$
 & $99.64$
\\
r3
 & $1.00$
 & $1.91$
 & $109.70$
 & $137.64$
\\
r4
 & $1.00$
 & $1.74$
 & $132.68$
 & $176.72$
\\
r5
 & $1.01$
 & $1.51$
 & $161.86$
 & $217.73$
\\
\hline
Average
 & $1.01$
 & $1.77$
 & $110.13$
 & $141.25$
\\
\hline\hline
\multicolumn{5}{c}{\textbf{randomaccess}}\\ \hline
& OD & RM & ROSEL & ROSES \\ \hline
r1
 & $0.98$
 & $1.02$
 & $11.20$
 & $12.26$
\\
r2
 & $0.98$
 & $1.02$
 & $15.80$
 & $18.91$
\\
r3
 & $0.96$
 & $1.02$
 & $20.83$
 & $24.25$
\\
r4
 & $0.96$
 & $1.00$
 & $28.16$
 & $34.28$
\\
r5
 & $0.95$
 & $0.99$
 & $42.22$
 & $56.51$
\\
\hline
Average
 & $0.96$
 & $1.01$
 & $23.64$
 & $29.24$
\\
\hline\hline
\end{tabular}
\end{center}
\end{table}

\revision{In contrast, the decoding complexity difference with respect to reference decoding is marginal. In the case of ROSES, the mean decoding time is $1.8\%$ and $1.9\%$ larger than the reference decoding time, for all-intra and randomaccess coding, respectively. }

\section{Conclusion}
\label{sec:concl}
This paper presented a novel rate-distortion optimized signal extrapolation algorithm for efficient compression of point clouds. The method selects few basis functions for the reconstruction of the occupied regions while simultaneously reducing the bitrate. Evaluations showed that for all-intra coding, more than $4\%$ of bitrate can be saved for the geometry map and more than $6\%$ of bitrate can be saved for the texture map. 

In future work, we plan to optimize the implementation of ROSES for a lower processing complexity \revision{by code optimization or restricting ROSES to a subset of selected coding modes, e.g., only the finally determined coding mode. Furthermore, it is planned} to develop an optimized rate model for the randomaccess configuration. Finally, the ROSES shall be implemented in VVC encoding. 

\section*{Acknowledgment}
This work was partly funded by the Deutsche Forschungsgemeinschaft (DFG, German Research Foundation) – SFB 1483 – Project-ID 442419336, EmpkinS.

\bibliographystyle{IEEEtran}

\begin{IEEEbiography}[{\includegraphics[width=1in,clip,keepaspectratio]{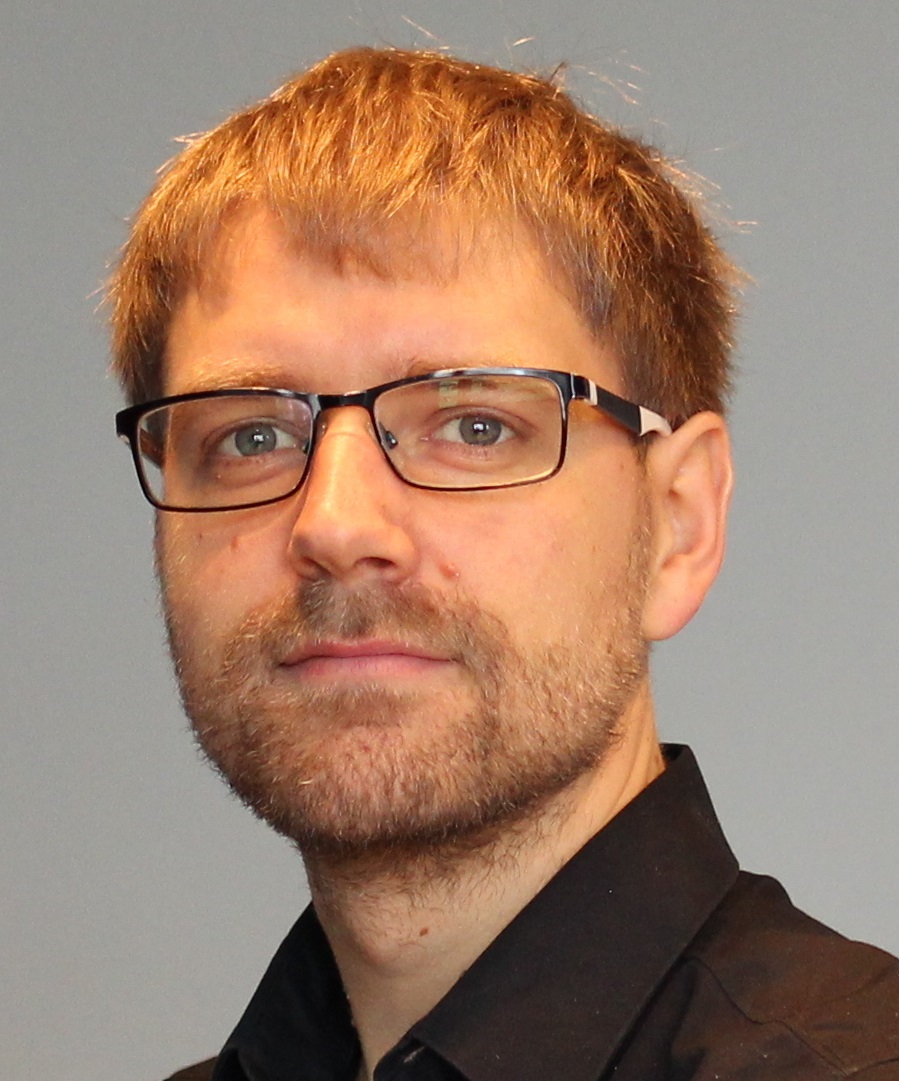}}]{Christian Herglotz} (SM'14-M'18)
 received the Dipl.-Ing. in electrical engineering and information technology in 2011 and the Dipl.-Wirt. Ing. in business administration and economics in 2012, both from Rheinisch-Westf\"alische Technische Hochschule (RWTH) Aachen University, Germany. Since 2012, he was a Research Scientist with the Chair of Multimedia Communications and Signal Processing, Friedrich-Alexander University Erlangen-N\"urnberg (FAU), Germany, where he received his Dr.-Ing. degree in 2017.

In 2018 and 2019, he worked as a PostDoc-Fellow at \'Ecole de technologie sup\'erieure (\'ETS) in collaboration with Summit Tech Multimedia, Montr\'eal, Canada on energy efficient VR technologies. Since 2019, he is with Friedrich-Alexander University Erlangen-N\"urnberg as a senior scientist. His current research interests include energy efficient video communications and visual signal compression. 

Since 2020, he is with the Visual Signal Processing and Communications Technical Committee of the IEEE Circuits and Systems Society. He served as a guest editor for the open journal on circuits and systems (OJCAS).  
\end{IEEEbiography}

\begin{IEEEbiography}[{\includegraphics[width=1in,clip,keepaspectratio]{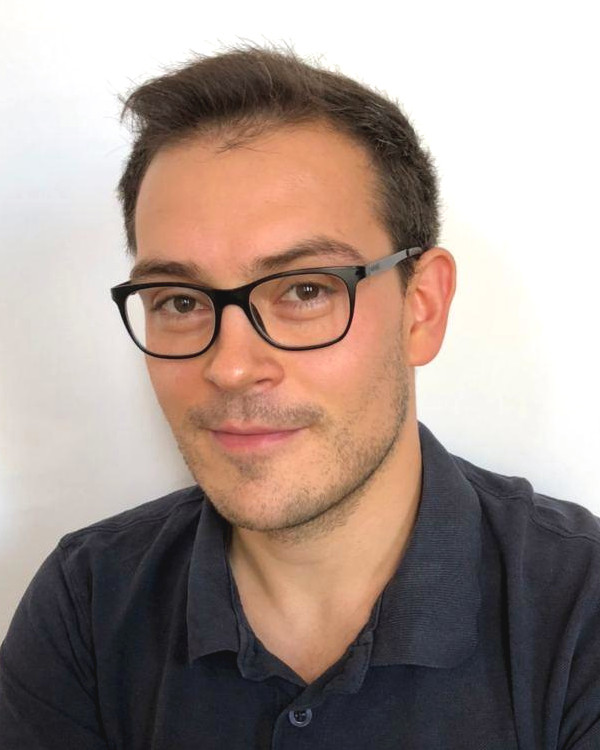}}]{Nils Genser} (SM'17)
received the master’s degree in information and communication technology from Friedrich-Alexander University Erlangen-Nürnberg (FAU), Germany, in 2016.

During his masters, he worked on image reconstruction and conducted his thesis at the Chair of Multimedia Communications and Signal Processing, FAU, where he continued as researcher. Among other things, he received a best master thesis award at FAU and a best student paper award at IWSSIP, in 2017.

His research interests include image and video signal processing, reconstruction, and coding. Moreover, he conducts research on spectral imaging, especially on regression-based and deep-learning-based registration and reconstruction algorithms.
\end{IEEEbiography}

\begin{IEEEbiography}[{\includegraphics[width=1in,clip,keepaspectratio]{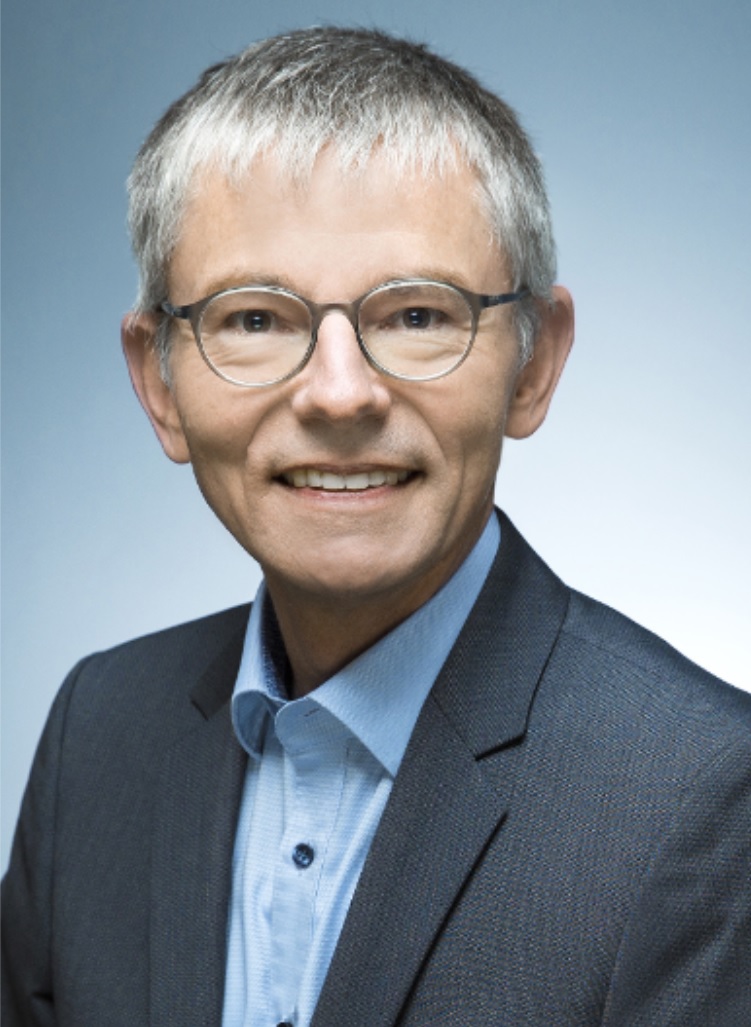}}]{Andr\'e Kaup} (M’96-SM’99-F’13)
received the Dipl.-Ing. and Dr.-Ing. degrees in electrical engineering from RWTH Aachen University, Aachen, Germany, in 1989 and 1995, respectively.

He joined Siemens Corporate Technology, Munich, Germany, in 1995, and became the Head of the Mobile Applications and Services Group in 1999.
Since 2001, he has been a Full Professor and the
Head of the Chair of Multimedia Communications
and Signal Processing at Friedrich-Alexander University Erlangen-N\"urnberg (FAU), Germany. From
2005 to 2007 he was Vice Speaker of the DFG Collaborative Research Center 603. From 2015 to 2017, he served as the Head of the Department of Electrical
Engineering and the Vice Dean of the Faculty of Engineering at FAU. He has authored around 400 journal and conference papers and has over 120 patents granted or pending. His research interests include image and video signal
processing and coding, and multimedia communication.

Dr. Kaup is a member of the IEEE Multimedia Signal Processing Technical Committee and the Scientific Advisory Board of the German VDE/ITG. In 2018, he was elected as a Full Member with the Bavarian Academy of
Sciences. He was a Siemens Inventor of the Year 1998 and received the
1999 ITG Award and several IEEE Best Paper Awards. His group won the
Grand Video Compression Challenge from the Picture Coding Symposium
2013. The Faculty of Engineering with FAU and the State of Bavaria honored him with Teaching Awards, in 2015 and 2020, respectively. He served as an
Associate Editor for IEEE Transactions on Circuits and Systems for Video Technology. He was a Guest Editor for IEEE Journal of Selected Topics in
Signal Processing.
\end{IEEEbiography}


\end{document}